\documentclass[12pt]{article}
\usepackage{enumerate}
 \usepackage{amsmath,amssymb,bm}
\usepackage{graphicx,wrapfig,lipsum,multirow,subfigure}
\usepackage{natbib}
\usepackage{ dsfont }
\usepackage{booktabs}
\usepackage{authblk}
\usepackage{color}
\usepackage{amsthm}
\usepackage{geometry}
\usepackage{comment}
\usepackage{url}
\usepackage{ wasysym }
\usepackage[table]{xcolor}
\usepackage{mathtools}
\usepackage[onehalfspacing]{setspace}
\newcommand{\uu}{\underline}
\definecolor{lavender}{RGB}{180, 52, 235}

\title{A Comparison of Model-Free Phase I Dose Escalation Designs for Dual-Agent Combination Therapies}
\author{Helen Yvette Barnett$^1$, Matthew George$^{2,3}$, Donia Skanji$^4$, Gaelle Saint-Hilary$^4$, Thomas Jaki$^{1,2}$, Pavel Mozgunov$^1$\\$^1$ MRC Biostatistics Unit, University of Cambridge\\
$^2$ Department of Mathematics and Statistics, Lancaster University\\
$^3$ Phastar\\
$^4$ Servier}

\begin{document}

\maketitle

\begin{abstract}
It is increasingly common for therapies in oncology to be given in combination. In some cases, patients can benefit from the interaction between two drugs, although often at the risk of higher toxicity. A large number of designs to conduct phase I trials in this setting are available, where the objective is to select the maximum tolerated dose combination (MTC). Recently, a number of model-free (also called model-assisted) designs have provoked interest, providing several practical advantages over the more conventional approaches of rule-based or model-based designs. In this paper, we demonstrate a novel calibration procedure for model-free designs to determine their most desirable parameters. Under the calibration procedure, we compare the behaviour of model-free designs to a model-based approach in a comprehensive simulation study, covering a number of clinically plausible scenarios. It is found that model-free designs are competitive with the model-based design in terms of the proportion of correct selections of the MTC. However, there are a number of scenarios in which model-free designs offer a safer alternative. This is also illustrated in the application of the designs to a case study using data from a phase I oncology trial. 
\end{abstract}


\clearpage
\section{Introduction}
The aim of phase I clinical trials investigating a single therapy is to find the highest dose that can be administered whilst ensuring that patients are at a low risk of serious side effects. To offer patients a higher chance of successful treatment, there is willingness to accept a dose that leads to more toxic responses, commonly labelled as dose-limiting toxicities (DLTs). The highest dose for which the treatment has a prespecified probability of leading to a toxic outcome is called the maximum tolerated dose (MTD). In an analysis of over 400,000 clinical trials conducted between 2000 and 2015                                                                                                                  \citep{wong2019estimation}, it was found that 57.6\% of all phase I oncology trials successfully progressed to phase~II. It was found that in 73\% of trials excluding oncology, treatments were successful in moving to phase~II, thus demonstrating the importance of successful dose-finding methods in oncology, where drugs are clearly harder to develop.

In this work, we consider phase I oncology trials in which a combination of two therapies are investigated. Here the objective is to identify a maximum tolerated dose combination (MTC), the highest dose combination with a probability of toxicity at the target. Phase I oncology trials in this dual-agent setting have recently provoked notable interest \citep{wong2016changing}. In particular, it was found that immunotherapy, a targeted agent that stimulates the immune system to fight cancerous cells \citep{couzin2013cancer}, can provide benefit to patients when administered in combination with chemotherapy or another targeted agent \citep{sharma2015immune}. One difficulty in the dual-agent setting is that the order of toxicity is unknown for some combinations -- if the amount of one compound in the combination is increased while another is decreased, it is unknown whether the overall toxicity goes up or down.

A number of dose-finding methods for dual-agent combination phase I trials relaxing the monotonicity assumption on the order of some of the combinations have been proposed in the literature. They broadly belong to one of three categories; rule-based, model-based and model-free (also known as model-assisted) designs. Rule-based designs (e.g. 3+3+3 or extensions of this \cite{Hamberg2009}) rely on a number of prespecified rules to determine when a dose is escalated, de-escalated and chosen as the MTD. Model-based designs (e.g. six-parameter model \cite{thall2003dose}) model the relationship between dose and probability of toxicity through a parametric function. Through the course of a trial, parameter estimates are updated to better describe this relationship. The model-free designs~\citep[][]{abbas2020comparison,mozgunov2019information}, do not pre-specify any relationship between dose and toxicity, thus do not rely on any parametric assumptions in their search for the MTD. However, unlike rule-based designs, the decision process in which the dose can be escalated or de-escalated is assisted with a statistical model.

Despite numerous papers demonstrating flaws in rule-based designs and their performance in drug combination trials \citep{riviere2015competing, thall2003practical, conaway2019impact}, it was reported that less than 5\% of combination trials in oncology between 2011 and 2013 deviated from rule-based designs \citep{riviere2015designs}. It is perhaps the restrictions associated with model-based designs, such as difficulty of implementation or communication to clinicians, that have made these less commonly used in real trials. Recently, model-free designs have attracted attention due to their practicality \citep{wages2016practical}, although these have not yet been fully evaluated in the literature.

The objective of this work is to review four recently proposed model-free dose-finding designs for phase I dual-agent combination studies, namely, the Bayesian Optimal Interval  design~\citep[BOIN]{Lin2017boin}, the Keyboard design~\citep[KEY]{yan2017keyboard}, the surface-free design~\citep[SFD]{mozgunov2020surface}, and the product of independent beta probabilities design~\citep[PIPE]{mander2015product}. We evaluate their performance in an extensive simulation study. To compare the methods on equal grounds, we propose a calibration procedure that selects the parameters of each of the designs that maximise the proportion of correct selections (subject to a safety constraint). We compare the performance of these designs to the Bayesian Logistic Regression Model (BLRM), a model-based approach that uses a two-parameter logistic model for each compound~\citep{neuenschwander2015bayesian}, as well as a non-parametric optimal benchmark. We also evaluate the performance of each of the designs in a case study of neratinib and temsirolimus~\citep{Gandhi2014}, to highlight the differences between approaches in a real trial setting.

The rest of the paper continues as follows. We provide a review of model-free designs in Section~\ref{sec:model-free}, before using a novel method to calibrate the parameters of each design leading to good performance in Section~\ref{sec:cal}. We then present detailed results from our simulation study across a wide range of toxicity scenarios in Section~\ref{sec:results}, including a conventional model-based design for comparison. Each design is also applied to a real case study of a dose finding trial from combination therapies in Section~\ref{sec:example}. We finish with a discussion of our results and thoughts in Section~\ref{sec:discussion}. 

\section{Methodological Review}\label{sec:model-free}
In this section, we describe the dose escalation procedure for each of the four approaches in a general dose-finding trial. It is assumed patients enter the trial in cohorts, and the dose combination for the next cohort is assigned once the previous cohort's responses are available. We first define the admissible combinations for each design. These are the dose combinations that are allowable for assignment for the next cohort of patients based on the last tested combination. We then describe the details of the escalation procedure in each of the designs in the following setting. Consider a dual-agent trial with $I$ doses of drug A, denoted $d_1^A < \dots < d_I^A$ and $J$ doses of drug B, denoted $d_1^B < \dots < d_J^B$. Let $d_{ij}$ represent the combination of doses $d_i^A$ and $d_j^B$ for $i=1,\ldots,I$ and $j=1,\ldots,J$. The total number of patients who receive combination $d_{ij}$ and the number of those who experience a toxic response on $d_{ij}$ during the trial are denoted $n_{ij}$ and $y_{ij}$ respectively. The probability of toxic response at $d_{ij}$ is written as $\pi_{ij}$ and the target toxicity is denoted $\phi$.

\subsection{Admissible Combinations} \label{sec:admiss}
Before deciding on a dose for the next cohort, each design defines a set of combinations that are admissible; i.e. combinations that the next cohort could be allocated to. These are best illustrated with a diagram, Figure \ref{fig:neighbour}. Suppose we are at $d_{22}$ in Figure \ref{fig:neighbour}, indicated by the `$\newmoon$' symbol. Admissible combinations for the BOIN and KEY designs are the same combination or adjacent combinations to the current one, represented by the `$\fullmoon$' symbols. 

In addition to these combinations, the other designs we consider also allow for diagonal de-escalation, where the next cohort is administered a combination that is one dose level lower in each drug, and also allow for anti-diagonal escalation, meaning the next cohort receives a combination that is one dose level higher in one drug and one dose level lower in the other. These are depicted by the `$\ast$' symbols in Figure \ref{fig:neighbour}, where reaching $d_{11}$ requires diagonal de-escalation and reaching $d_{31}$ or $d_{13}$ requires anti-diagonal escalation. The rationale is that by enabling faster movement across the combination grid, the design can move to the MTC quickly, and de-escalate quickly if patients are treated at highly toxic combinations.

All designs prohibit diagonal escalation, where the next cohort receives a combination one dose level higher in each drug and no dose levels can be skipped. These non-admissible doses are shown by the '$\times$' in the red cells in Figure \ref{fig:neighbour}.
\begin{figure}[ht!]
\centering
\begin{tabular}{c|cccc}
$d_4^A$ & \cellcolor{red!70}$\times$ & \cellcolor{red!70}$\times$ & \cellcolor{red!70}$\times$ & \cellcolor{red!70}$\times$ \\ 
$d_3^A$ & $\ast$ & \fullmoon & \cellcolor{red!70}$\times$ & \cellcolor{red!70}$\times$ \\
$d_2^A$ & \fullmoon & \newmoon & \fullmoon & \cellcolor{red!70}$\times$ \\
$d_1^A$ & $\ast$ & \fullmoon & $\ast$ & \cellcolor{red!70}$\times$ \\
\hline 
& $d_1^B$ & $d_2^B$ & $d_3^B$ & $d_4^B$
\end{tabular}
\caption{Illustration of the admissible combinations for each design. The `$\newmoon$' symbol illustrates the current dose combination, and the symbols `$\fullmoon$' and `$\ast$' represent the possible combinations the next cohort could receive for different designs. \label{fig:neighbour}}
\end{figure}

\subsection{The BOIN Design} \label{sec:boin}
The Bayesian Optimal Interval (BOIN) design \citep{Lin2017boin} uses the intuitive estimator $\hat{\pi}_{ij} = y_{ij}/n_{ij}$ for the probability of toxicity at combination $d_{ij}$, so that $\hat{\pi}_{ij}$ is the proportion of observed toxic responses on $d_{ij}$ across the whole trial. The estimator $\hat{\pi}_{ij}$ only updates after patient responses are observed on $d_{ij}$, and is then used to guide dose escalation. This escalation process is defined by pre-specified values of $\lambda_e, \lambda_d$ to which $\hat{\pi}_{ij}$ is compared after each cohort. Values of $\lambda_e < \phi$ and $\lambda_d > \phi$ are chosen to locally minimise the chance of incorrect escalation and de-escalation decisions during a trial, and are calculated using constants $\phi_1$ and $\phi_2$. Whilst $\phi$ is the target toxicity, $\phi_1$ is the highest toxicity probability deemed sub-therapeutic and $\phi_2$ is the lowest toxicity probability deemed overly toxic. These can be specified by the clinicians. $\lambda_e $ and $\lambda_d $ are defined in Equation~(\ref{eq:boin}):
\begin{equation}
\lambda_e = \frac{\log \left( \frac{1-\phi_1}{1-\phi} \right) }{\log \left( \frac{\phi(1-\phi_1)}{\phi_1(1-\phi)} \right)} \;\; \text{and} \;\; \lambda_d = \frac{\log \left( \frac{1-\phi}{1-\phi_2} \right)}{\log \left( \frac{\phi_2(1-\phi)}{\phi(1-\phi_2)} \right)}. \label{eq:boin}
\end{equation}
Both $\lambda_e$ and $\lambda_d$ are invariant to $d_{ij}$, $n_{ij}$ and $y_{ij}$, so that optimising these parameters depends only on constants $\phi$, $\phi_1$ and $\phi_2$. After defining $\lambda_e$ and $\lambda_d$, the rules for the dose-finding procedure are as follows:
\begin{itemize}
\item If $\hat{\pi}_{ij} \leq \lambda_e$, the next combination is chosen from $\mathcal{A}_E = \left\lbrace d_{(i+1)j}, d_{i(j+1)} \right\rbrace$. 
\item If $\hat{\pi}_{ij} \geq \lambda_d$, the next combination is chosen from $\mathcal{A}_D = \left\lbrace d_{(i-1)j}, d_{i(j-1)} \right\rbrace$.
\item Otherwise, $\lambda_e < \hat{\pi}_{ij} < \lambda_d$ and the next combination is the same. 
\end{itemize}
In this way, dose skipping, diagonal escalation and diagonal de-escalation are prohibited -- see Section \ref{sec:admiss} for more details. If the next combination is to be chosen from an empty $\mathcal{A}_E$ or $\mathcal{A}_D$ (for example the current combination is the highest in both doses and the design chooses to escalate), then the next cohort receives the same combination. The design assumes each patient response is independent, $y_{ij} \sim \text{Binomial}(n_{ij},\pi_{ij})$ and assigns a vague Beta(1,1) prior distribution to each $\pi_{ij}$, giving the posterior distribution for $\pi_{ij}$ as
\begin{equation}
    \pi_{ij} | n_{ij}, y_{ij} \sim \text{Beta}(y_{ij} + 1, n_{ij} - y_{ij} + 1). \label{eq:boin.mod}
\end{equation}
To choose between combinations in the chosen set, the BOIN design computes the posterior probability $\mathbb{P}(\pi_{ij} \in (\lambda_e, \lambda_d) | n_{ij}, y_{ij})$. The combination maximising this probability is administered to the next cohort. For combinations yet to be tested, calculating this probability is based on the vague prior distribution only. In the event of ties, which is always the case when multiple potential combinations are yet to be administered, the next combination is selected at random from the chosen set. Note that no toxicity information is borrowed between the combinations under this model as the combinations are treated independently.  

The design uses an overdosing criterion stating that a combination, and any that are more toxic under monotonicity, satisfying $\mathbb{P}(\pi_{ij} > \phi | n_{ij}, y_{ij}) \geq \epsilon_{BOIN}$ for some overdosing probability threshold $0 < \epsilon_{BOIN} \leq 1$, cannot be administered to the next cohort. For the BOIN design, if $d_{ij}$ satisfies this condition, dose $d_{ij}$ and higher combinations are eliminated from the trial, and the dose maximizing $\mathbb{P}(\pi_{ij} \in (\lambda_e, \lambda_d) | n_{ij}, y_{ij})$ within $\mathcal{A}_D$ is chosen for the next cohort. If combination $d_{11}$ satisfies the overdosing criterion, the trial is terminated earlier for safety.

After all patients are treated, estimates of each $\pi_{ij}$ are calculated via matrix isotonic regression \citep{dykstra1982algorithm}. The simple technique guarantees that estimates of $\pi_{ij}$ at higher combinations are at least as high as estimates of $\pi_{ij}$ at lower combinations, which follows the assumption of monotonicity. The MTC is selected as the combination with estimated $\pi_{ij}$ closest to $\phi$ via isotonic regression \citep{dykstra1982algorithm}. 

\subsection{The Keyboard Design} \label{sec:key}
The Keyboard design (KEY) \citep{yan2017keyboard} is very similar to the BOIN design, defining an interval about the target toxicity $\phi$, denoted $I_{target} = (\phi - \Delta_1, \phi + \Delta_2)$, for constants $\Delta_1, \Delta_2 > 0$, which can be chosen by the clinicians. A combination with estimated toxicity probability within this interval is said to have acceptable toxicity. The design then divides the (0,1) space into ``keys'', defined as intervals $I_t$ of equal length $\Delta_1 + \Delta_2$ (allowing for shorter keys at either end of (0,1)) for $t=1,\ldots,T$, where $T$ is the number of keys. The interval $I_{target}$ is fixed pre-trial, chosen to minimise the chance of incorrect escalation and de-escalation decisions. 

The KEY design assigns a vague Beta(1,1) prior distribution to each $\pi_{ij}$, and assumes that the number of toxic responses follows a binomial distribution, $y_{ij} \sim \text{Binomial}(n_{ij}, \pi_{ij})$. The posterior distribution for each $\pi_{ij}$ is computed as in Equation \ref{eq:boin.mod}. Again, this means there is no borrowing of toxicity information across combinations. The design then identifies the key $I_t$ that is most likely to contain $\pi_{ij}$, labelled $I_{max}$,
\begin{equation}
    I_{max} = \underset{I_t: t\in(1,\ldots,T)}{\operatorname{argmax}} \; \mathbb{P}(\pi_{ij} \in I_t | n_{ij}, y_{ij}). \label{eq:key}
\end{equation}
Once the key $I_{max}$ is identified, escalation and de-escalation decisions happen as follows: 
\begin{itemize}
\item If $I_{max} < I_{target}$, the next combination is chosen from $\mathcal{A}_{E} = \left\lbrace d_{(i+1)j}, d_{i(j+1)} \right\rbrace$.
\item If $I_{max} > I_{target}$, the next combination is chosen from $\mathcal{A}_{D} = \left\lbrace d_{(i-1)j}, d_{i(j-1)} \right\rbrace$.
\item If $I_{max} = I_{target}$, the next combination is the same.
\end{itemize}
To choose between combinations in $\mathcal{A}_E$ (or $\mathcal{A}_D$), the design computes the posterior probability $\mathbb{P}(\pi_{ij} \in I_{target} | n_{ij}, y_{ij})$ for all combinations in $\mathcal{A}_E$ (or $\mathcal{A}_D$). The combination maximising this probability is administered to the next cohort. The remainder of the escalation process and the selection of the MTC is analogous with the BOIN design, with an identical overdosing rule using $\epsilon_{KEY}$ and the MTC chosen via isotonic regression \citep{dykstra1982algorithm}. 

Both the BOIN and KEY designs model dose combinations independently, however in the following two designs, the connections between the dose combinations are also taken into account.

\subsection{The Surface-Free Design} \label{sec:sfd}
The surface-free design (SFD) \cite{mozgunov2020surface} does not restrict the MTC search to a parametric surface and does not require the order of toxicity between combinations to be known. The main idea is to parametrise ratios between toxicity probabilities for different combinations, defining $\theta = 1 - \pi_{11}$, and $\theta_i=\frac{1-\pi_{i,j}}{1-\pi_{i-1,j}}$. Then $\theta$ is the probability of a patient having no toxic response on the lowest dose combination and $\theta_i$ denotes the ratio between the probability of a patient having no toxic response on dose combinations $d_{ij}$ and $d_{(i-1)j}$ for $j=2,\ldots,J$ and $i=2,\ldots,I$. Similarly, $\tau_j=\frac{1-\pi_{i,j}}{1-\pi_{i,j-1}}$ is defined as the ratio between the probability of a patient having no toxic response on $d_{ij}$ and $d_{i(j-1)}$ for $j=2,\ldots,J$ and $i=2,\ldots,I$. Thus, the probability of toxicity for each combination $d_{ij}$ is 
\begin{equation}
    \pi_{ij} = 1-\theta \theta_2 \ldots \theta_i \tau_2 \ldots \tau_j. \label{eq:sfd}
\end{equation}
Due to monotonicity, each ratio $\theta_i, \tau_j \in(0,1)$ and the SFD assigns each of these ratios an independent Beta prior distribution. The hyper-parameters of the prior distributions can be chosen to match the clinicians' prior mean estimates of toxicity probability on each combination and effective sample sizes.

After each cohort, the SFD updates the posterior means for ratios $\theta,\theta_2,\ldots,\theta_I,\tau_2,\ldots,\tau_J$ using Bayes theorem, which can be related back to $\pi_{ij}$ through Equation~(\ref{eq:sfd}) to give estimates of the toxicity probabilities. In this way, the SFD is borrowing information across various drug combinations previously collected in the trial to make an informed decision on escalation. Additionally, the continual multiplication of Beta random variables implies that $\pi_{ij}$ for higher combinations has higher variance, allowing for more cautious escalation at higher combinations.  Considering all neighbouring combinations apart from the one higher in both doses, the next combination is chosen as the one with estimated $\pi_{ij}$ closest to $\phi$. An overdosing criterion prohibits any combination from being administered if $\mathbb{P}(\pi_{ij} > \phi | n_{ij}, y_{ij}) \geq \epsilon_{SFD}$ for some $\epsilon_{SFD} > 0$, and the trial is terminated if this is satisfied for $d_{11}$. 

Once all patients have been treated, the MTC is selected as the combination with toxicity probability closest to $\phi$. Note that the SFD design is more computationally intensive than the other model-free designs as MCMC methods are required to sample from the posterior distribution.

\subsection{The PIPE Design} \label{sec:pipe}
The PIPE design \citep{mander2015product} differs from other model-free designs in that it was originally proposed to find the MTC contour, labelled $MTC_\phi$. This is a line partitioning the combination space into safe and overly toxic combinations. Those below the contour are believed to have toxicity probability less than target toxicity $\phi$, whilst those above are believed to have toxicity probability greater than $\phi$. 

Assuming the $\pi_{ij}$ are independent, they are assigned a Beta prior distribution, $\pi_{ij}\sim\text{Beta}(a_{ij},b_{ij})$ for hyper-parameters $a_{ij}$ and $b_{ij}$, for $i=1,\ldots,I$ and $j=1,\ldots,J$. Priors can be prespecified if knowledge on the toxicity of combinations is available. Assuming each patient is independent such that $y_{ij} \sim \text{Binomial}(n_{ij},\pi_{ij}) \; \forall i,j$, the posterior for $\pi_{ij}$ can be written as
\begin{equation}
    \pi_{ij} | n_{ij}, y_{ij} \sim \text{Beta}(y_{ij} + a_{ij}, n_{ij} - y_{ij} + b_{ij}). \label{eq:pipe.mod}
\end{equation}
The posterior distribution is only updated after a cohort of patients is treated on the corresponding combination, but the $MTC_{\phi}$ is re-estimated regardless of which combination was tested. The monotonicity assumption means that the PIPE design needs only to consider contours satisfying this property, limiting the number of possible contours to $\binom{I+J}{I}$.

Each contour can be represented by a binary matrix, where entries are 0 or 1 depending on whether estimates of the toxicity probability for a combination are below or above the contour respectively. Let $\vartheta$ be the set of all monotonic contours for an $I \times J$ dose combination space and define $C_s \in \vartheta$ as the binary matrix representing the contour $s=1,\ldots,\binom{I+J}{I}$. 

To estimate the $MTC_\phi$ given the current data, the design calculates the posterior probability of each toxicity probability being less than or equal to $\phi$, that is 
\begin{equation}
p_{ij}(\phi | n_{ij}, y_{ij}) = \mathbb{P}(\pi_{ij} \leq \phi | n_{ij}, y_{ij}, a_{ij}, b_{ij}), \label{eq:pipe.prob}
\end{equation}
where the right-hand side of Equation~(\ref{eq:pipe.prob}) is equal to the cumulative distribution function of a Beta distribution. Equation~(\ref{eq:pipe.general}) gives the general formula for calculating the probability that the $MTC_{\phi}$ is defined by the matrix $C_s$:
\begin{equation}
\mathbb{P}(MTC_{\phi} = C_s | n_{ij}, y_{ij}) = \prod_{i=1}^I \prod_{j=1}^J \left\lbrace 1 - p_{ij}(\phi | n_{ij}, y_{ij}) \right\rbrace ^{C_s[i,j]} p_{ij}(\phi | n_{ij}, y_{ij}) ^{1-C_s[i,j]}, \label{eq:pipe.general}
\end{equation} 
where $[i,j]$ represents the entry in the $i$th row and $j$th column of the binary matrix $C_s$. The contour maximising Equation~(\ref{eq:pipe.general}) is the contour most likely to be the $MTC_{\phi}$ given the current data. This contour then assists the escalation process by identifying the combinations closest to it, before the design selects one of these for the next cohort based on a weighted randomisation procedure. This involves weighting each combination by the inverse of their sample size, with the rationale being varied experimentation around the $MTC_{\phi}$. Escalation continues in this way until all patients are treated, at which point all combinations closest from below the $MTC_{\phi}$ are recommended for phase II.


The design uses an overdosing rule which considers the expected probability of $d_{ij}$ being above the most probable $MTC_{\phi}$ averaged over all monotonic contours. This is written as
\begin{equation*}
    q_{ij} = \sum_{C_s \in \vartheta} C_s[i,j] \mathbb{P}(MTC_{\phi} = C_s | \bm{Y}^{(m)}),
\end{equation*}
and $d_{ij}$ cannot be administered to the next cohort if $q_{ij} \geq \epsilon_{PIPE}$ for some $\epsilon_{PIPE} > 0$. A trial is terminated if combination $d_{11}$ satisfies this condition.

The PIPE design can recommend multiple combinations for phase II, as it recommends all combinations closest from below its $MTC_{\phi}$. For consistency across designs, in our implementation we ensure only one combination is recommended as the MTC. Therefore for each recommended combination, we find the posterior mean probability of toxicity, which can be calculated using the posterior distributions in Equation~(\ref{eq:pipe.prob}). The combination with posterior mean closest to $\phi$ is selected as the MTC, only choosing a combination at random in the event of a tie.

\section{Calibration of Designs} \label{sec:cal}

Model-based and model-free designs based on a Bayesian framework give clinicians more control over their performance. The PIPE design, the SFD and most model-based designs allow for knowledge on the toxicity of each drug from monotherapy trials to be incorporated into the design through their prior distributions. As the BOIN and KEY designs assign vague priors to the toxicity probabilities, their behaviour is primarily determined by the pre-defined intervals guiding escalation. Although it is in theory possible to incorporate historical data through the prior in the BOIN and KEY designs, for the purpose of this comparison, it would defeat the purpose of a design with all escalation boundaries pre-specified at the design stage for ease of implementation.

In this comparison study, the purpose of the calibration procedure is to give all designs a set-up which leads to consistently high proportions of selections of combinations with toxicity probability close to $\phi$ in all scenarios. To achieve this, we calibrate each design using a novel two-stage approach. The first stage of the calibration is concerned with choosing values for hyper-parameters that give a good performance in selecting the MTC without considering safety. The second stage then focusses on safety, calibrating the overdose rule taking into account not only good performance in terms of correct selections, but also the number of patients who are treated at unsafe doses.

This approach employs a grid search over hyper-parameter or interval values (depending on the design), each stage involving running simulations over four clinically plausible scenarios and determining which values lead to superior performance. We refer to the priors resulting in superior performance across the four scenarios as operational priors. Although, for the purposes of this work, they will serve as a way of a fairer comparison between the Bayesian designs, addressing the challenge of ensuring that the same amount of prior information is used for each design, the obtained operational priors can be also applicable in the practical case where no reliable prior information about the compound is available. 

To evaluate which design inputs lead to superior performance in recommending the MTC in the first stage, the proportion of correct selections (PCS) is examined in each of the four scenarios. That is, the proportion of trials in which a design selects any combination with a true toxicity probability of exactly 0.30. To summarise overall performance across these four scenarios, the geometric mean PCS is considered. Suppose $x_1, \ldots, x_N$ represent the PCS in $N$ scenarios. The geometric mean, $\left( \prod_{i=1}^N x_i \right)^{1/N}$, is used instead of the arithmetic mean because it has the useful property of penalising cases in which PCS are more dispersed across scenarios. The design with priors or intervals resulting in highest geometric mean PCS across the four scenarios will be the design variant we choose. For the remainder of this section, the mean will refer to the geometric mean. We note that during the first stage of the calibration procedure, no overdosing rules are included, meaning no trials are to be stopped before all patients have been recruited, because we choose to calibrate the parameter controlling the overdosing rule in the separate second stage. Once the first stage of calibration is complete, this will lead to the selection of intervals for the BOIN and KEY designs, and operational priors for the PIPE, SFD and BLRM designs. 

\begin{figure}[h!]
\centering
\subfigure[BOIN: Patients treated at overly toxic doses]{\includegraphics[width=0.42\textwidth]{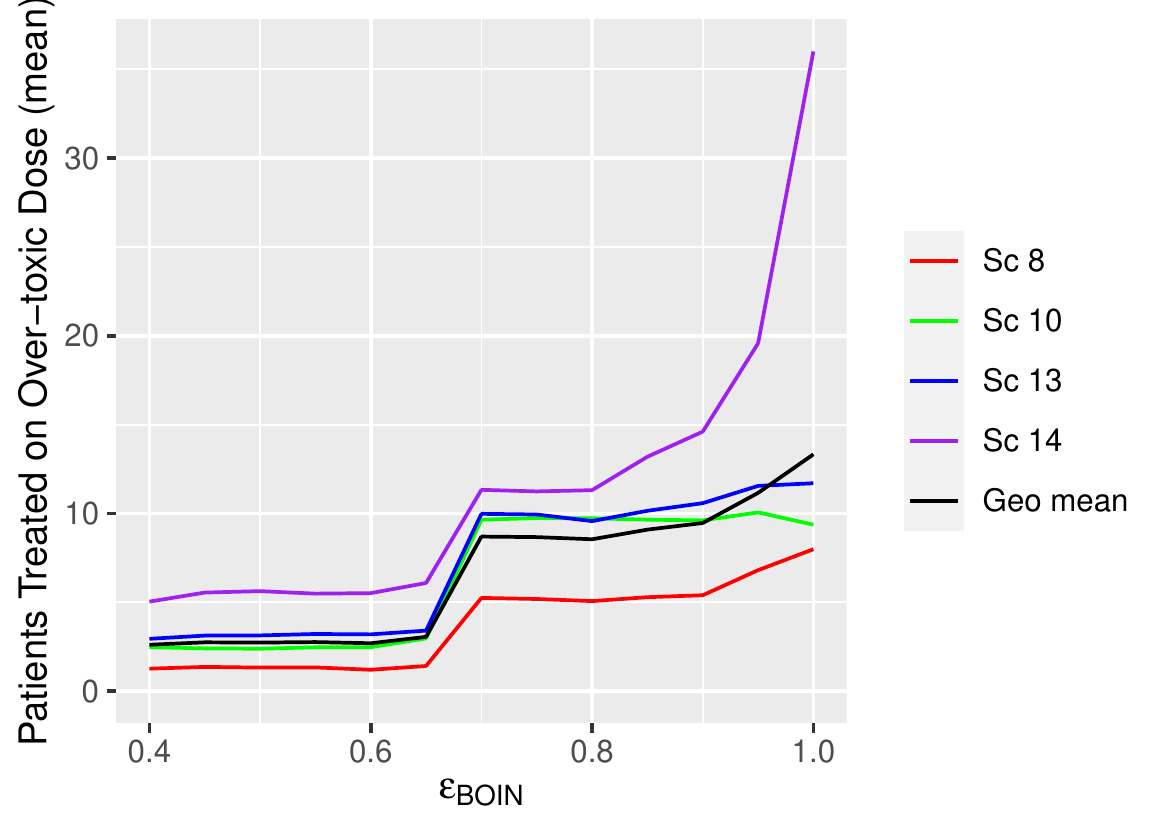}}
\hspace{0.05\textwidth}
\subfigure[BOIN: Proportion of correct selections]{\includegraphics[width=0.42\textwidth]{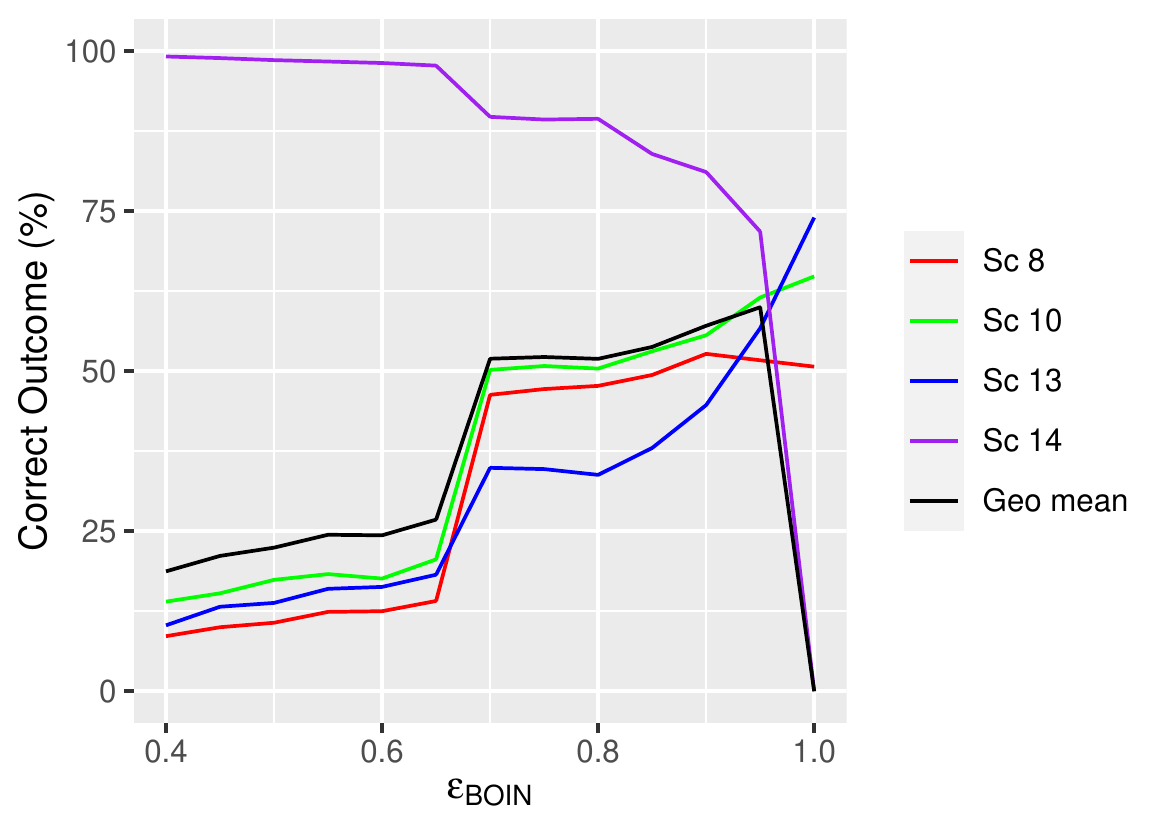}}

\subfigure[KEY: Patients treated at overly toxic doses]{\includegraphics[width=0.42\textwidth]{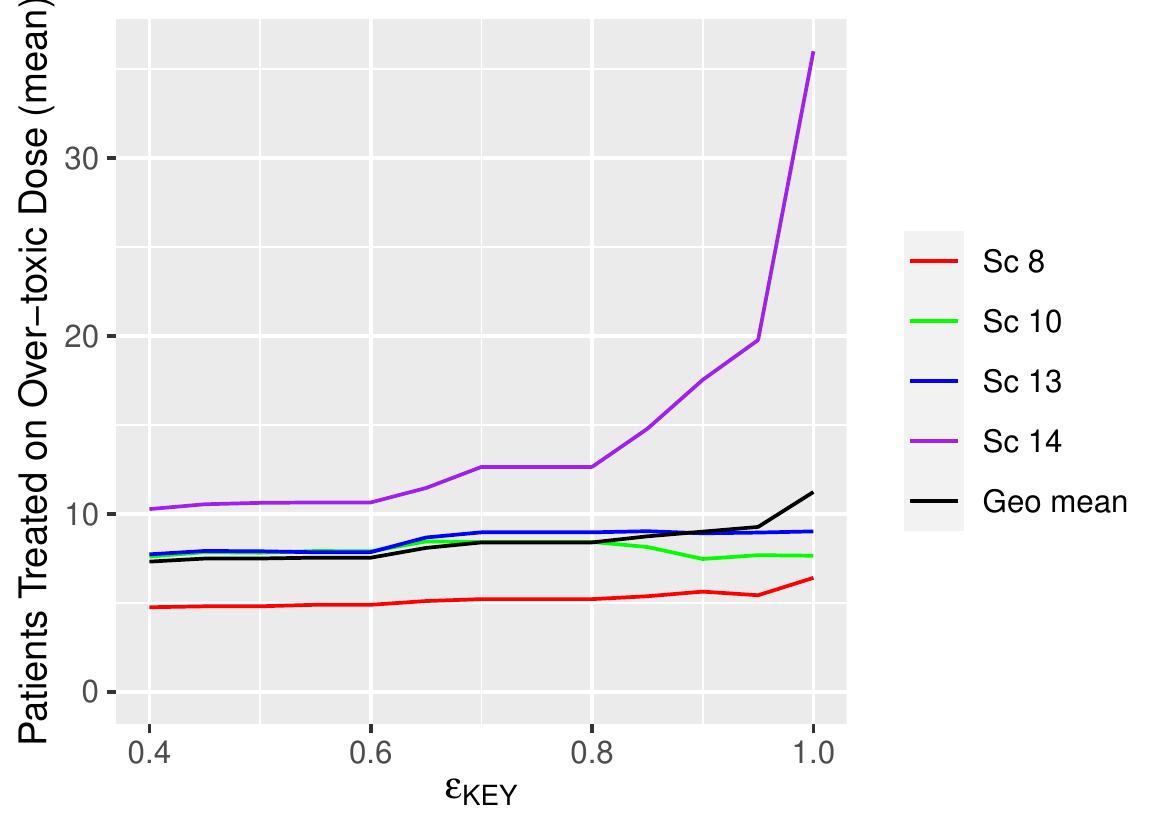}}
\hspace{0.05\textwidth}
\subfigure[KEY: Proportion of correct selections]{\includegraphics[width=0.42\textwidth]{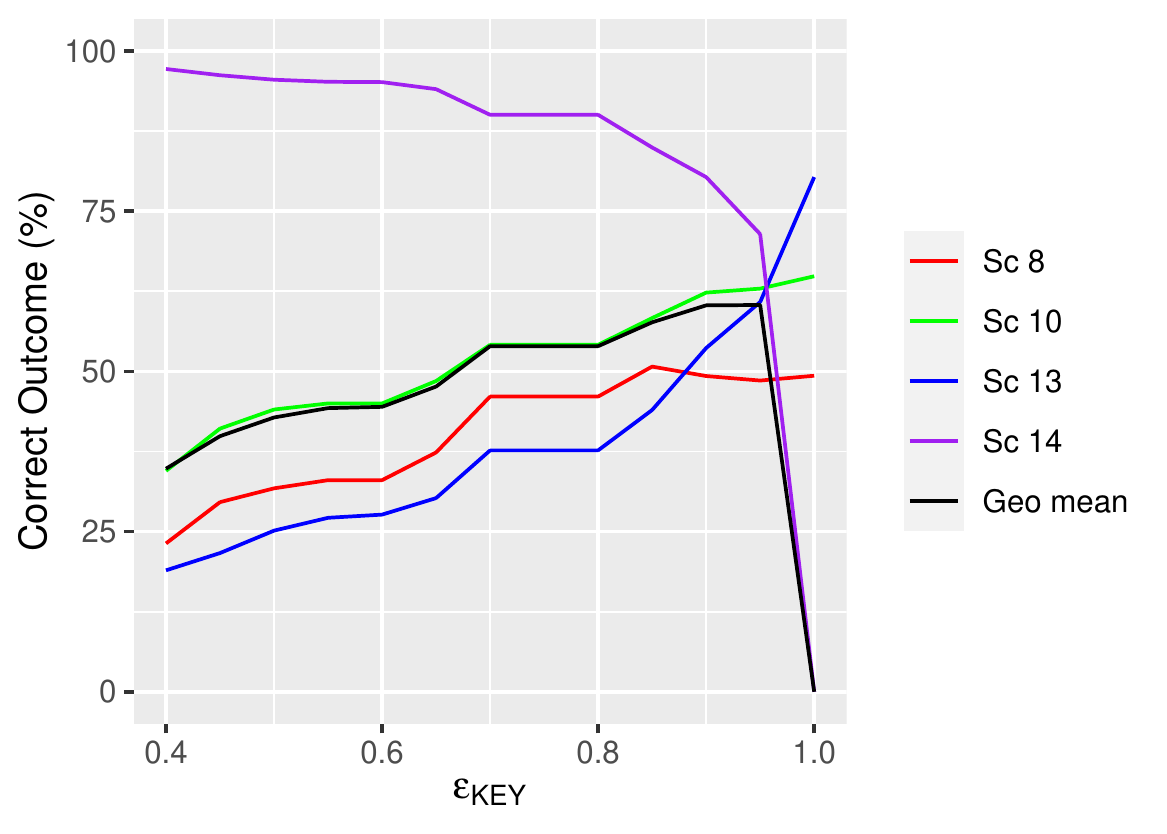}}

\subfigure[Surface-Free: Patients treated at overly toxic doses]{\includegraphics[width=0.42\textwidth]{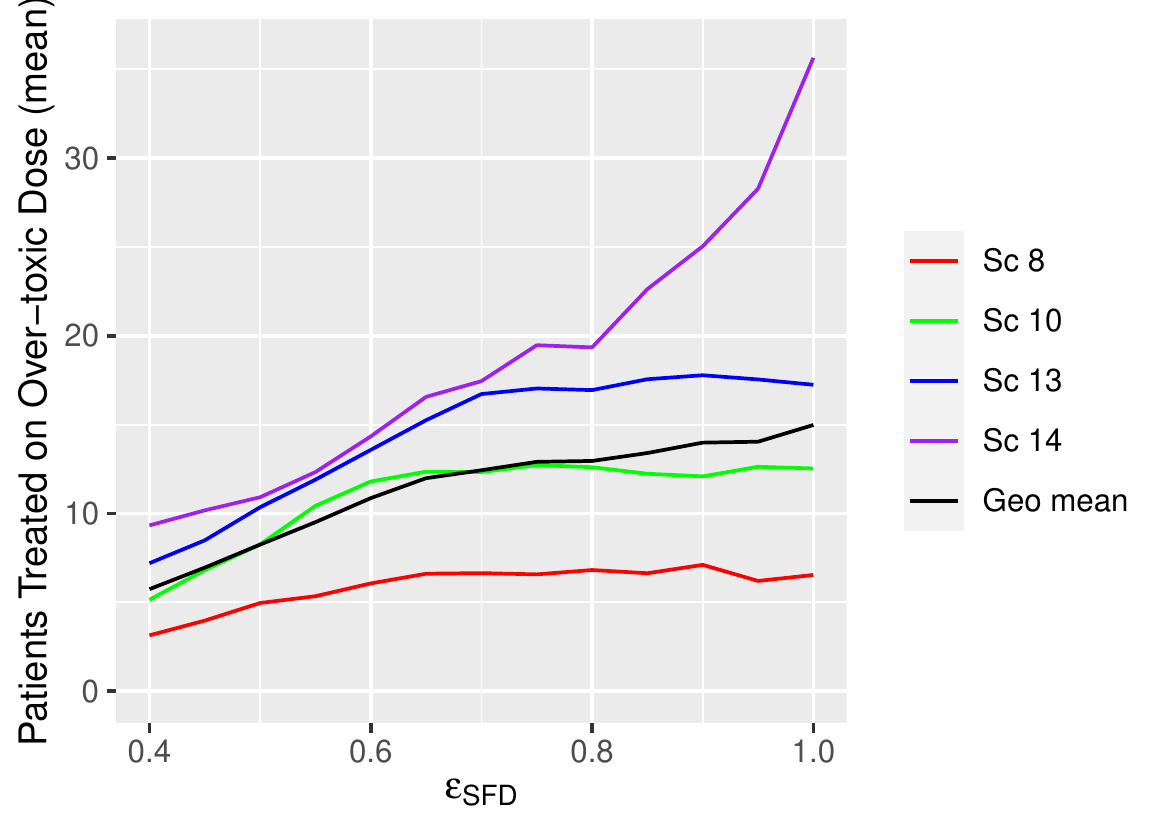}}
\hspace{0.05\textwidth}
\subfigure[Surface-Free: Proportion of correct selections]{\includegraphics[width=0.42\textwidth]{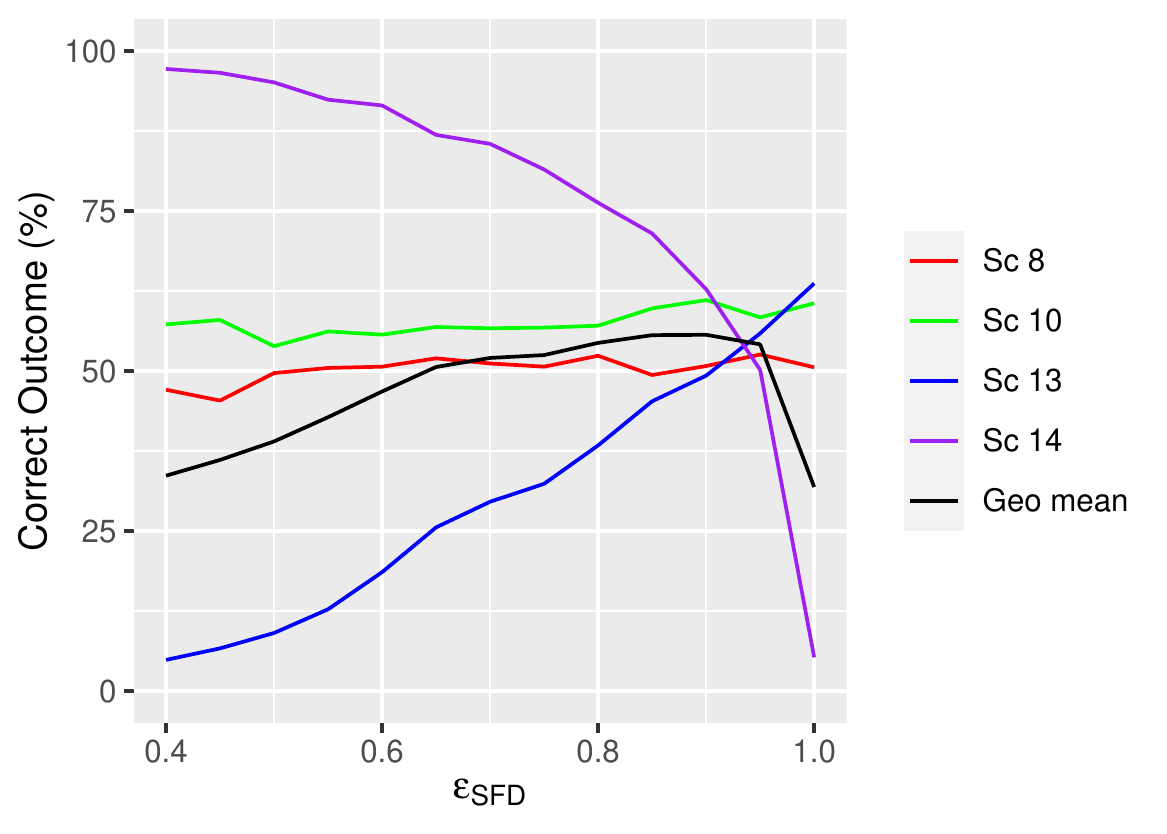}}

\subfigure[PIPE: Patients treated at overly toxic doses]{\includegraphics[width=0.42\textwidth]{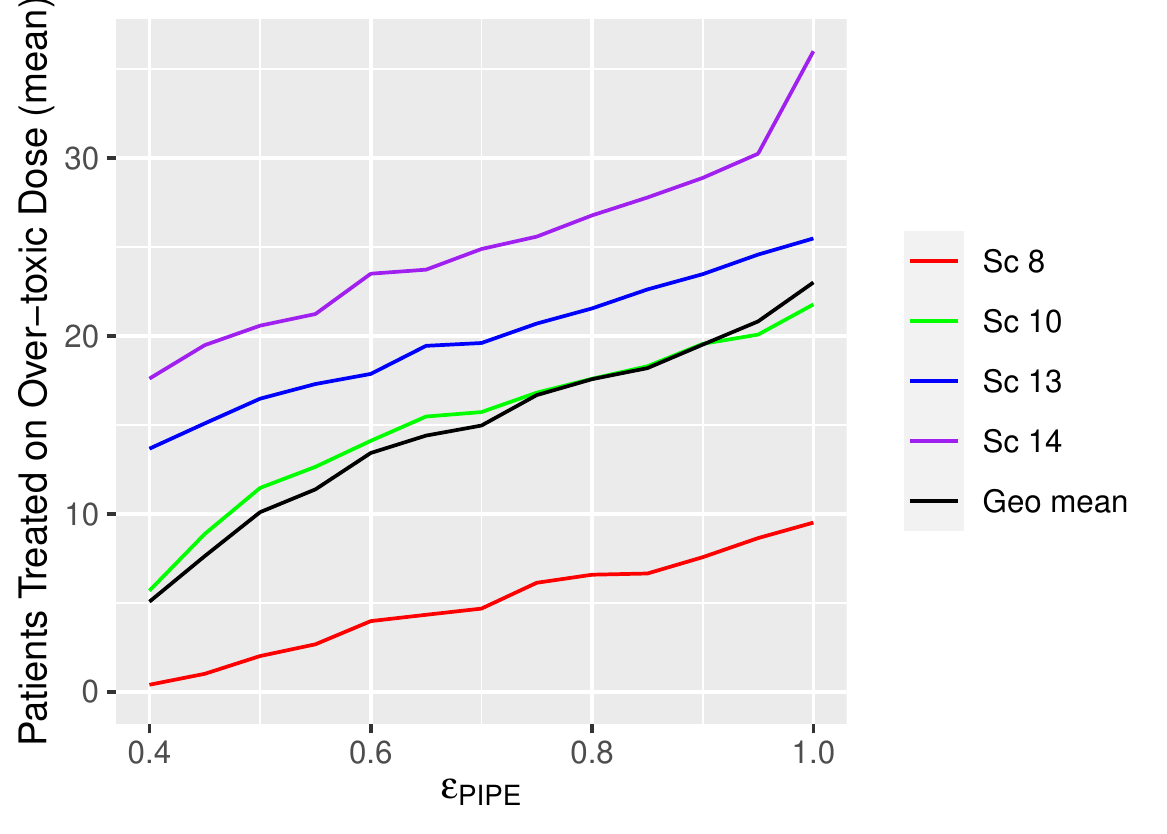}}
\hspace{0.05\textwidth}
\subfigure[PIPE: Proportion of correct selections]{\includegraphics[width=0.42\textwidth]{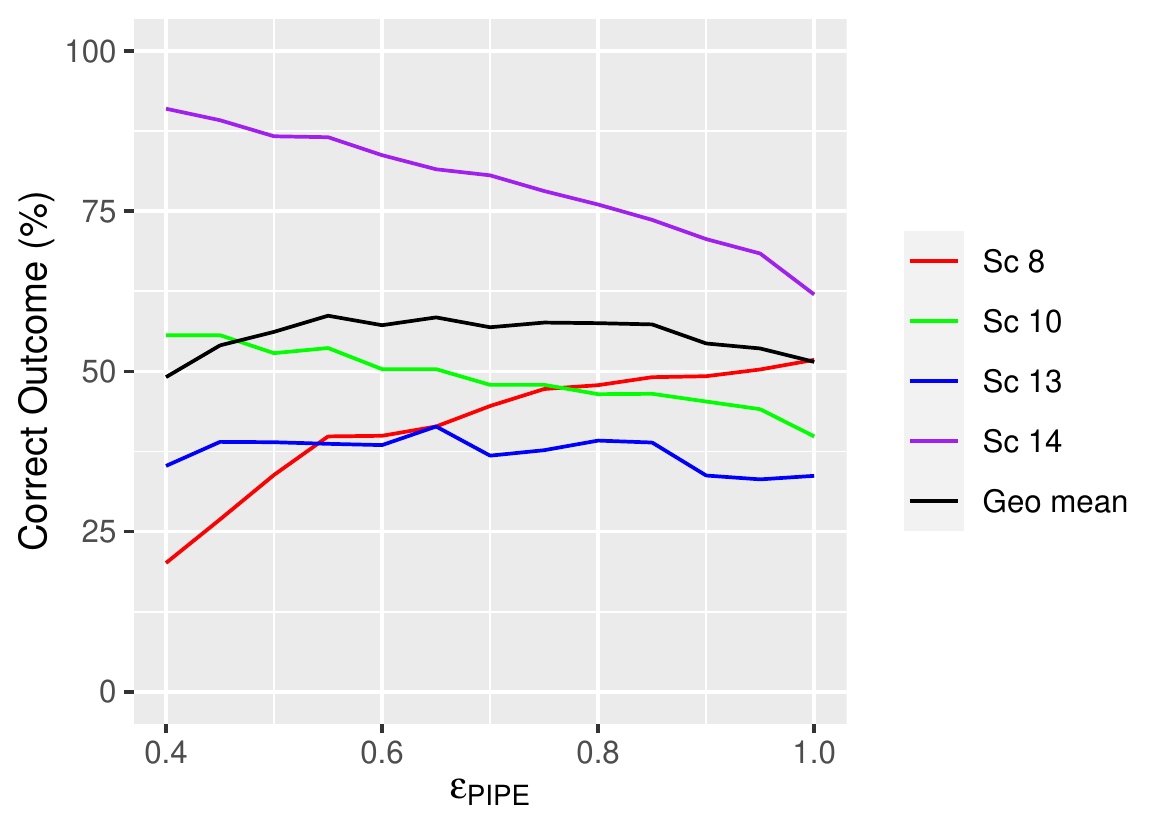}}

 \caption{Calibration of $\epsilon$ for the four designs}
  \label{fig:CAL_eps}
\end{figure}

The second stage of the calibration procedure is for $\epsilon$, the parameter regulating the overdosing rule in each model-free design. Calibration of $\epsilon$ involves decreasing its value starting from 1, and observing the proportion of correct outcomes in the chosen scenarios. As selecting overly toxic combinations is more of an ethical concern, as a general rule we take as a starting point the highest value of $\epsilon$ resulting in at least 85\% of trials recommending no combinations when considering an overly toxic scenario. We acknowledge this proportion may differ in practice depending on the clinicians' judgement. It is important to note that the interpretation of $\epsilon$ differs between designs because of the construction of each overdosing rule, and should be accounted for when communicating with clinicians. This is reflected by subscripts for the individual designs in the following specifications. The second stage of the calibration procedure for each design is illustrated in Figure~\ref{fig:CAL_eps}.

\subsection{Setting}

Each design is calibrated in the same setting that is then explored in the simulation study (see Section~\ref{sec:sim_setting}), representative of a phase I trial in oncology. There are two drugs with three dose levels each, which results in nine combinations, and the first cohort is treated at the lowest combination. The objective is to select a single combination as the MTC with true toxicity probability $\phi = 0.30$. The sample size is 36 patients for which are recruited in cohorts of three patients. All combination-toxicity scenarios are presented in Table \ref{tab:scenarios}. However, four scenarios are chosen to explore noticeably different clinical cases, in which the number and location of the MTCs vary, whilst restricting the number of scenarios makes the procedure computationally feasible.

In stage 1 of the calibration procedure, Scenarios 1, 8, 10 and 13 are chosen. Scenarios 1 and 13 are chosen to represent the extremes: when the highest combination is the only true MTC and all others are safe, and when the lowest combination is the only true MTC and all others are overly toxic, respectively. Scenario 8 covers situations in which most combinations are safe but true MTCs do not lie on the same diagonal. Scenario 10 captures the case where most combinations are overly toxic and true MTCs lie on the same diagonal. Note that we often refer to the set of combinations in a scenario as the combination grid.

In stage 2, simulations are run for each design over Scenarios 8, 10, 13 and 14 for different values of $\epsilon$. In Scenarios 8, 10 and 13, the PCS is as previously defined, whilst in the unsafe Scenario 14 we consider the PCS as the proportion of trials in which no combinations are selected. We refer to selecting no combinations in Scenario 14 as the `correct outcome'.

\subsection{Calibrating the BOIN Design} \label{sec:boincal}

To guide dose escalation, the BOIN design relies on the interval $(\lambda_e, \lambda_d)$ around the target toxicity. Interval boundaries $\lambda_e$ and $\lambda_d$ are a function of $\phi$, $\phi_1$ and $\phi_2$, where $\phi_1=a_1\phi$ and $\phi_2=a_2\phi$ for constants $a_1<1, a_2>1$. To calibrate the design, we run 4000 simulations for each scenario for pairs $(a_1, a_2)$ from the sets $a_1 = \{0.85, 0.80, \ldots, 0.40\} $ and $a_2 = \{ 1.15, 1.20, \ldots, 1.60\}$, resulting in a total of 100 pairs. As constants $a_1$ and $a_2$ deviate further from 1, the interval becomes wider, thus the design will choose to escalate and de-escalate on fewer occasions.

The optimal values are found to be $a_1=0.65$ and $a_2=1.4$, which substituting into Equation~(\ref{eq:boin}), we generate the interval boundaries $\lambda_e$ and $\lambda_d$ to give the interval (0.245, 0.359) to guide dose escalation. This interval implies that escalation occurs if $\pi_{ij}$ is below 0.245, de-escalation occurs if $\pi_{ij}$ is above 0.359, else the combination remains the same.

In the second stage of calibration, we find that as $\epsilon_{\mbox{\tiny{BOIN}}}$ decreases, the design benefits more in Scenario 14, where the proportion of trials in which no combinations are recommended increases (see Figure~\ref{fig:CAL_eps}). For $\epsilon_{\mbox{\tiny{BOIN}}} \leq 0.84$, over 85\% of trials recommend no combinations in Scenario 14. The trade-off in the other scenarios with this $\epsilon_{\mbox{\tiny{BOIN}}}$ value is that PCS increases steeply when $\epsilon_{\mbox{\tiny{BOIN}}}$ increases, as well as the number of patients treated on overly toxic doses increasing. Therefore $\epsilon_{\mbox{\tiny{BOIN}}}=0.84$ is chosen. 

\subsection{Calibrating the Keyboard Design} \label{sec:keycal}
Using a similar method to BOIN, we first calibrate the parameters which define the interval for KEY. The interval $I_{target} = (b_1, b_2)$ guides escalation entirely so is an important component of the design. We run 4000 simulations across each scenario for pairs $(b_1, b_2)$ from the sets $b_1 = \{0.27, 0.25, \ldots, 0.19\} $ and $b_2 = \{0.33, 0.35, \ldots, 0.41\}$, resulting in a total of 25 pairs. Mean PCS are displayed Figure 2 in the online supplementary materials, indicating that interval (0.21, 0.39) yields the highest mean PCS, which differs from the recommendation of (0.25, 0.35) in the original paper \citep{yan2017keyboard}. As explained in Section \ref{sec:key}, this means escalation occurs only if the posterior probability $\mathbb{P}(\pi_{ij} \in (0.03, 0.21) | n_{ij}, y_{ij})$ is higher than $\mathbb{P}(\pi_{ij} \in (0.21, 0.39) | n_{ij}, y_{ij})$.

In the second stage of calibration, we find that as $\epsilon_{\mbox{\tiny{KEY}}}$ decreases, the design benefits more in Scenario 14, where the proportion of trials in which no combinations are recommended increases. Choosing $\epsilon_{\mbox{\tiny{KEY}}}=0.84$ leads to approximately 85\% of trials correctly selecting no combinations in Scenario 14, as shown in Figure~\ref{fig:CAL_eps}, in line with the value obtained for the BOIN design.

\subsection{Calibrating the Surface-Free Design} \label{sec:sfdcal}
The SFD assigns Beta priors to each of its parameters; the ratios between toxicity probabilities. In this setting, there are five ratios ($\theta$, $\theta_2$, $\theta_3$, $\tau_2$ \& $\tau_3$ defined in Section~\ref{sec:sfd}) to parametrise, meaning a total of 10 hyper-parameters for the beta priors must be defined for the operational priors. Instead of specifying these directly, we specify a prior mean and prior effective sample size for each ratio, which can be used to calculate the corresponding hyper-parameters. To make the calibration task computationally feasible, we assume that all prior mean ratios, $m$, are equal (meaning the increase in dose corresponds to the same proportion increase in toxicity) and all effective sample sizes for each ratio, $s_{\mbox{\tiny{SFD}}}$, are equal. Thus we only need to calibrate pairs of $m$ and $s$, which we choose from sets $m = \{0.95, 0.925, \ldots, 0.85 \}$ and $s_{\mbox{\tiny{SFD}}} = \{1, 2, \ldots, 5 \}$.

For each pair, we run 500 simulations (which is lower than other model-free designs due to the computational demands of the design) and examine the mean PCS across the four scenarios. Our results in Figure 4 in the online supplementary materials show that the mean PCS is highest for $m=0.875$ and $s_{\mbox{\tiny{SFD}}}=4$. This is equivalent to every ratio being assigned the prior distribution $\text{Beta}(3.5, 0.5)$, and corresponds to mean prior toxicity probabilities on $d_{11}$ and $d_{33}$ of 0.125 and 0.487 respectively.

For the calibration of $\epsilon_{\mbox{\tiny{SFD}}}$, in Figure~\ref{fig:CAL_eps}, we found that $\epsilon_{\mbox{\tiny{SFD}}}=0.65$ resulted in at least 85\% of trials selecting no combinations in Scenario 14. There is evidence to suggest that increasing or decreasing $\epsilon_{\mbox{\tiny{SFD}}}$ not only has a sizeable effect on the PCS in Scenario 13, but also the number of patients treated at unsafe doses, demonstrating the design is highly sensitive to changes in its overdosing rule.

\subsection{Calibrating the PIPE Design} \label{sec:pipecal}
Similar to the SFD, the PIPE designs assigns beta priors to each $\pi_{ij}$. A prior mean and prior sample size for each $\pi_{ij}$ are specified, giving a total of 18 values to specify from which the hyper-parameters for the beta priors can be calculated. To make calibration feasible, we assume that prior sample size $s_{\mbox{\tiny{PIPE}}}$  is equal for each combination and to set the prior means, we divide the grid of combinations into five diagonal segments, with toxicity increasing as we move through each segment. In this way, the design follows the monotonicity assumption. To assign a toxicity to each combination, we specify the toxicity of the lowest combination, $\rho$, and the size of the increments in toxicity between each segment, $\delta$. In the illustration in Figure \ref{fig:pipecal}, we have chosen $\rho=0.05$ and $\delta=0.025$ to construct the grid.
\begin{figure}[h!]
    \centering
    \begin{tabular}{|cc|ccc|}
        \hline
        \multicolumn{2}{|c|}{\multirow{2}{*}{$\searrow$}} & \multicolumn{3}{c|}{Drug B} \\
        \multicolumn{2}{|c|}{} & $d_1^B$ & $d_2^B$ & $d_3^B$ \\ 
        \hline
        \multirow{3}{*}{\rotatebox[origin=c]{90}{~Drug A}} & $d_1^A$ & \cellcolor{red!0}0.05 & \cellcolor{red!10}0.075 & \cellcolor{red!30}0.10 \\
        & $d_2^A$ & \cellcolor{red!10}0.075 & \cellcolor{red!30}0.10 & \cellcolor{red!60}0.125 \\ 
        & $d_3^A$ & \cellcolor{red!30}0.10 & \cellcolor{red!60}0.125 & \cellcolor{red!100}0.15 \\ 
        \hline
    \end{tabular}
    \caption{Constructing prior mean toxicity probabilities when calibrating the PIPE design. \label{fig:pipecal}}
\end{figure}

Our approach involves calibrating three parameters simultaneously to create operational priors, and are chosen from the sets $\rho = \{0.025, 0.05, 0.075, 0.10\}$, $\delta = \{0.025, 0.05, 0.075, 0.10\}$ and $s_{\mbox{\tiny{PIPE}}} = \{1/72, 1/36, 1/18, 1/9\}$. For each triple, we run 2000 simulations in each of the four scenarios, which is fewer than for the BOIN and KEY designs due to the minor increase in computational expense. We provide one grid in Figure 5 in the online supplementary materials to account for mean PCS on each $s_{\mbox{\tiny{PIPE}}}$ value. The triple $s_{\mbox{\tiny{PIPE}}}=1/18$, $\rho=0.05$ and $\delta=0.025$ leads to the highest mean PCS, although we observe that there were many triples that resulted in similar values. We note our choice of prior sample size, $s_{\mbox{\tiny{PIPE}}}=1/18$, only differs to the recommendation of 1/9 in the original paper \citep{mander2015product}. For prior sample sizes $s_{\mbox{\tiny{PIPE}}} \leq 1/18$, the design is found to be robust. Mean PCS only varies between 37\% and 40\%, suggesting that a number of operational priors could lead to consistently high PCS.

For the second stage of the calibration, the value of $\epsilon_{\mbox{\tiny{PIPE}}}$ is varied as shown in Figure~\ref{fig:CAL_eps}, and $\epsilon_{\mbox{\tiny{PIPE}}}=0.50$ is chosen as it provides at least an 85\% chance of correctly recommending no combinations in Scenario 14, as well as balancing the number of patients treated at unsafe doses in the four considered scenarios.

\section{Simulation Study} \label{sec:results}
In this section we describe the setting for the simulation study before presenting the results, including a comparison to a model-based approach and a non-parametric optimal benchmark.
\subsection{Setting} \label{sec:sim_setting}
In order to compare the discussed designs, we conduct a simulation study, performing 2000 simulations of each of the 15 scenarios depicted in Table~\ref{tab:scenarios} for all five designs.  As before, the objective is to select a single combination as the MTC with true toxicity probability $\phi = 0.30$. Any combination with probability of toxicity greater than 0.33 is labelled as overly toxic, and any combination with probability of toxicity in the interval [0.16,0.33] is labelled as acceptable. In this section, the mean refers to the arithmetic mean. All simulations are carried out using \texttt{R}~\citep{R}, with code provided in the online supplementary materials.
 
In general, the number of overly toxic combinations available for selection increases as we move through Scenarios 1 to 14. Scenario 1 has a single MTC which is the highest combination available. Scenarios 3 and 4 contain very few overly toxic combinations and have MTCs on the edge of the grid. Scenario 5 is similar to these, except its only MTC is located in the centre of the grid. In Scenarios 2, 6, 7, 8, 9 and 10, there are multiple combinations to explore which have toxicity probability $\phi$. In particular, Scenarios 8 and 9 aim to investigate design behaviour when underlying MTCs are not on the same diagonal. Scenarios 11, 12 and 13 represent settings in which most combinations are overly toxic, meaning designs should avoid combinations away from $d_{11}$. Scenario 14 is of importance because all of its combinations are overly toxic, making the trial very unethical. In this instance, the only correct outcome is to recommend no combination for phase II. Scenario 15 represents a situation where all combinations are true MTCs, and is used to monitor escalation behaviour when combinations are safe and increasing the dose of either drug does not affect toxicity.

In order to accentuate the differences in the designs, we do not implement any accuracy or sufficient information rules, as these may mask some key elements of the designs. We focus on the operating characteristics of proportion of correct selections (PCS) and proportion of acceptable selections (PAS) as measures of accuracy, and proportion of overly toxic selections and the number of patients treated on unsafe dose combinations as measures of safety.

\subsection{A Model-Based Comparator}
To provide a comparison between model-free and model-based designs, we also consider a conventional model-based approach in our simulation study, the Bayesian Logistic Regression Model (BLRM) \citep{neuenschwander2015bayesian}. In this approach, the toxicity probability for each combination, $\pi_{ij}$, are modelled as in Equation \ref{eq:blrm.mod} for $i=1,\ldots,I$ and $j=1,\ldots,J$, where doses $d_i^A$ and $d_j^B$ are scaled by reference doses. Let $d_{ij}$ be combination of $d_i^A$ and $d_j^B$, while $n_{ij}$ and $y_{ij}$ are the number of patients and toxic responses on each combination respectively. Parameters $\alpha_1$ and $\beta_1$ describe the toxicity of drug A, $\alpha_2$ and $\beta_2$ describe the toxicity of drug B, and $\eta$ models the interaction between drugs. The five parameters are assigned normal prior distributions, and the likelihood is a product of Bernoulli densities, proportional to $\prod_{i=1}^I \prod_{j=1}^J \pi_{ij}^{y_{ij}} (1-\pi_{ij})^{n_{ij} - y_{ij}}$. After each cohort is observed, the joint posterior distribution is approximated using MCMC methods, and samples of each parameter are drawn from their full conditional distributions. Estimates of $\pi_{ij}$ are made by sampling parameters from their posteriors and substituting these along with the corresponding doses into Equation \ref{eq:blrm.mod}. Note that all parameters except $\eta$ are sampled on the log scale and then exponentiated since they must be positive.
\begin{equation}
    \pi_{ij}(\alpha_1, \alpha_2, \beta_1, \beta_2, \eta | d_i^A, d_j^B) = \frac{\left[\alpha_1 (d_i^A)^{\beta_1} + \alpha_2 (d_j^B)^{\beta_2} + \alpha_ 1\alpha_2 (d_i^A)^{\beta_1} (d_j^B)^{\beta_2}\right] \exp(\eta d_i^A d_j^B)}{1 + \left[\alpha_1 (d_i^A)^{\beta_1} + \alpha_2 (d_j^B)^{\beta_2} + \alpha_1 \alpha_2 (d_i^A)^{\beta_1} (d_j^B)^{\beta_2}\right] \exp(\eta d_i^A d_j^B)}. \label{eq:blrm.mod}
\end{equation}
The BLRM can only escalate to combinations satisfying the neighbourhood constraint and Escalation With Overdose Control (EWOC) principle. The neighbourhood constraint prevents escalation or de-escalation to any combination that is more than one dose level of either drug away, and also prevents escalation to a combination in which both dose levels are higher. For a trial with target toxicity $\phi=0.30$, the EWOC principle states that $d_{ij}$ can only be administered if $\mathbb{P}(\pi_{ij} > 0.33) < \epsilon_{\mbox{\tiny{BLRM}}}$. The combination maximising the probabilistic statement $\mathbb{P}(0.16 < \pi_{ij} < 0.33)$ is administered to the next cohort. If no combinations satisfy the two constraints, the trial is terminated. Once the sample size has been exhausted, the MTC is selected from combinations which have been experimented on with at least six patients, and is the one maximising $\mathbb{P}(0.16 < \pi_{ij} < 0.33)$. The BLRM requires dosing quantities for each drug to be specified, in all of the implementations of the BLRM, these doses are 100, 200 and 300mg for each drug. The same proposed calibration procedure as is applied to the other designs is applied to the BLRM, with details provided in the online supplementary materials.

\subsection{A Non-Parametric Optimal Benchmark Comparator}

While the primary goal of this work is to compare the performance of different model-free designs to each other, there is a risk that all methods might perform equally poorly on some scenarios. In this case, the comparison of the designs to each other would not identify why the poor performance is observed -- due to the challenging scenario or due to all designs having difficulties identifying a particular MTC. To provide context for the comparison of operating characteristics, we include the performance of the non-parametric benchmark for combination studies, a tool that provides an estimate for the upper bound on the PCS under the given combination-toxicity scenario~\citep{mozgunov2020benchmark,mozgunov2021benchmark}. The benchmark takes into account the ``difficulty'' of a scenario in terms of how close the toxicity risks for the combinations (under this scenario) are to the target level of 30\%, and also accounts for the unknown monotonic ordering in the combination setting. We refer the reader to the recent work by Mozgunov et al. \cite{mozgunov2021benchmark} for further technical details on the benchmark for combinations implementation. 

\subsection{Results}

\subsubsection{Proportions of Correct and Acceptable Selections}

\begin{figure}[ht!]
    \centering
    \includegraphics[width=0.99\textwidth]{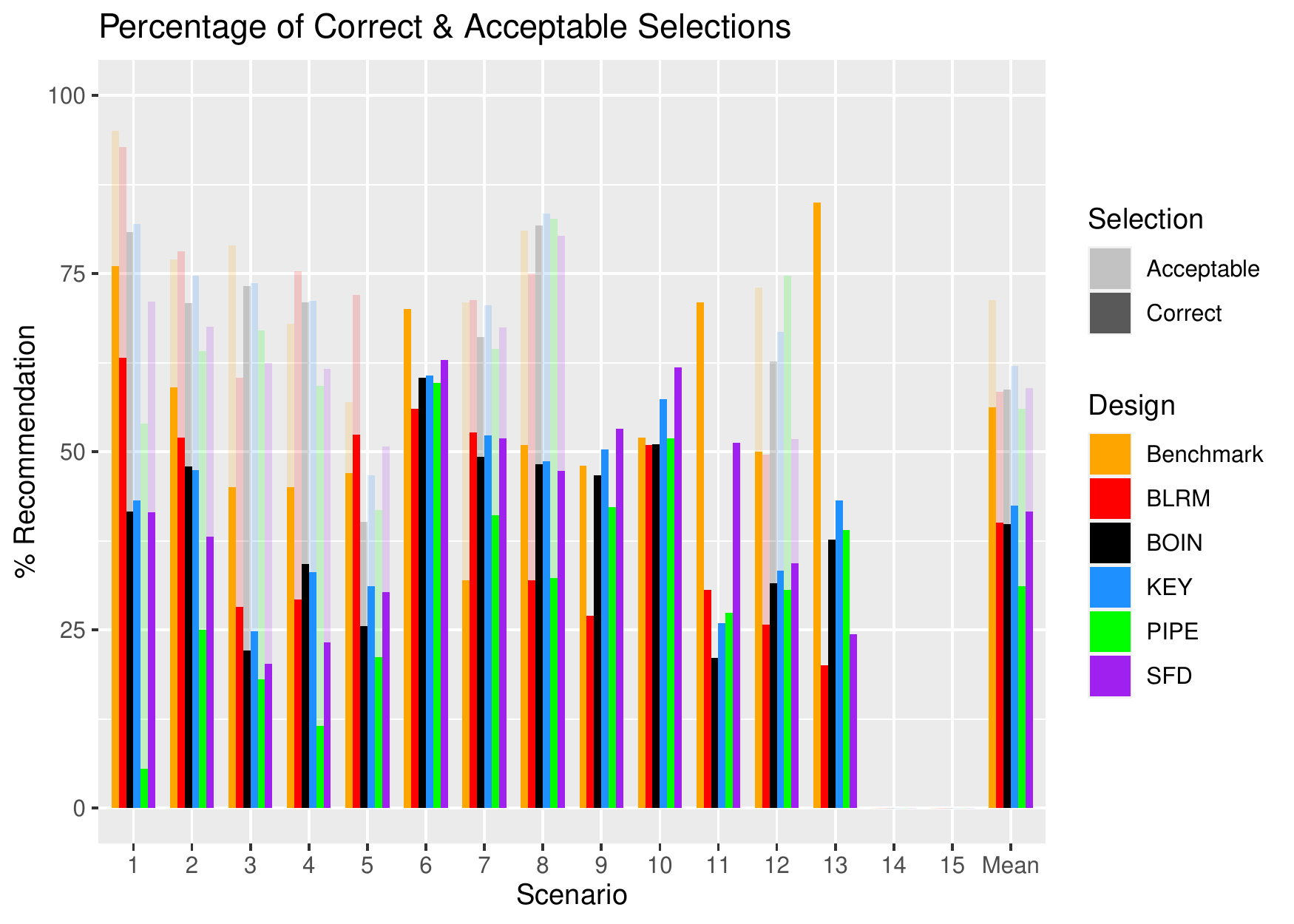}
    \caption{An illustration of the PCS and PAS for Scenarios 1-13 for each design. The solid bars measure the PCS and the more transparent bars measure the PAS. The rightmost group of bars show the means.}
    \label{fig:pcs}
\end{figure}

Figure \ref{fig:pcs} presents the summary of the operating characteristics of the considered designs in terms of the PCS and PAS (with the full set of results given in  the online supplementary materials). Scenarios 14 and 15 have been excluded as these have no true MTCs for the design to select. For scenarios in which the only acceptable combinations are also correct combinations (Scenarios 6, 9, 10, 11 and 13), the PCS and PAS are equal. The mean PCS across Scenarios 1-13 for the BLRM, BOIN, KEY, PIPE and SFD designs is 40.0\%, 39.8\%, 42.4\%, 31.2\% and 41.5\% respectively, whilst the mean PAS are 58.4\%, 58.7\%, 62.0\%, 56.0\% and 59.0\% respectively.

First of all, the benchmark reveals the differences in how challenging it is to identify the MTC in the considered scenarios: the PCS for the benchmark varies between approximately 35\% under Scenario 7 to more than 80\% under Scenario 13. As expected, the benchmark corresponds to the highest average PCS and PAS - 55\% and nearly 70\%, respectively. Similarly, under the majority of scenarios the benchmark corresponds to the highest PCS and PAS as it employs the concept of the complete information. The largest difference between the benchmark and other designs can be seen under Scenario~13. At the same time, there are scenarios under which the benchmark is outperformed by a competing design - this can be a sign of the design favouring particular combinations under the calibrated priors - for example under Scenario~7.

The variety of performances across the scenarios demonstrates the variability between the different designs in different settings. Considering the model-free designs, on average the KEY design has the highest proportion of both correct and acceptable selections, but is vastly outperformed in some scenarios by the SFD design. In six of the scenarios, the KEY has the highest PCS out of all the model-free designs, being superior in scenarios with few overly toxic combinations. However, for example in Scenario 11, where the MTC is the middle dose of drug A and lowest dose of drug B, the SFD outperforms the next best performing design by 20.6\%. The PIPE design shows poor performance in many scenarios, most notably in Scenario 1 where the PCS is 5.5\% and PAS is 54.0\%. A likely reason is that for the PIPE design, the choice of MTC must be below the MTC contour, and a scenario where the true MTC is the highest dose combination gives rise to underestimation since we cannot explore above the true MTC contour. In addition, the procedure discussed in Section~\ref{sec:pipe} to choose one MTC from the recommended set will make our results differ from those originally reported by Mander and Sweeting \cite{mander2015product}, where a `correct selection' was defined as the MTC being in the set of recommended doses.

When considering the BLRM as a comparator, we see that in many scenarios the BLRM outperforms the KEY. For example, in Scenario 1 where the MTC is the highest combination, the BLRM has PCS over 20\% higher than the next best performing design, the KEY. In fact, when including the BLRM in the comparison, the KEY is only the best performing design in one scenario, Scenario 8. The SFD does however outperform the BLRM in some cases, with the BLRM having the highest PCS in Scenarios 1, 2, 3, 5, and 7 and the SFD is the best performing in Scenarios 6, 9, 10, 11, and 12.

\subsubsection{Proportions of Overly Toxic Selections}
\begin{figure}[ht!]
    \centering
    \includegraphics[width=0.99\textwidth]{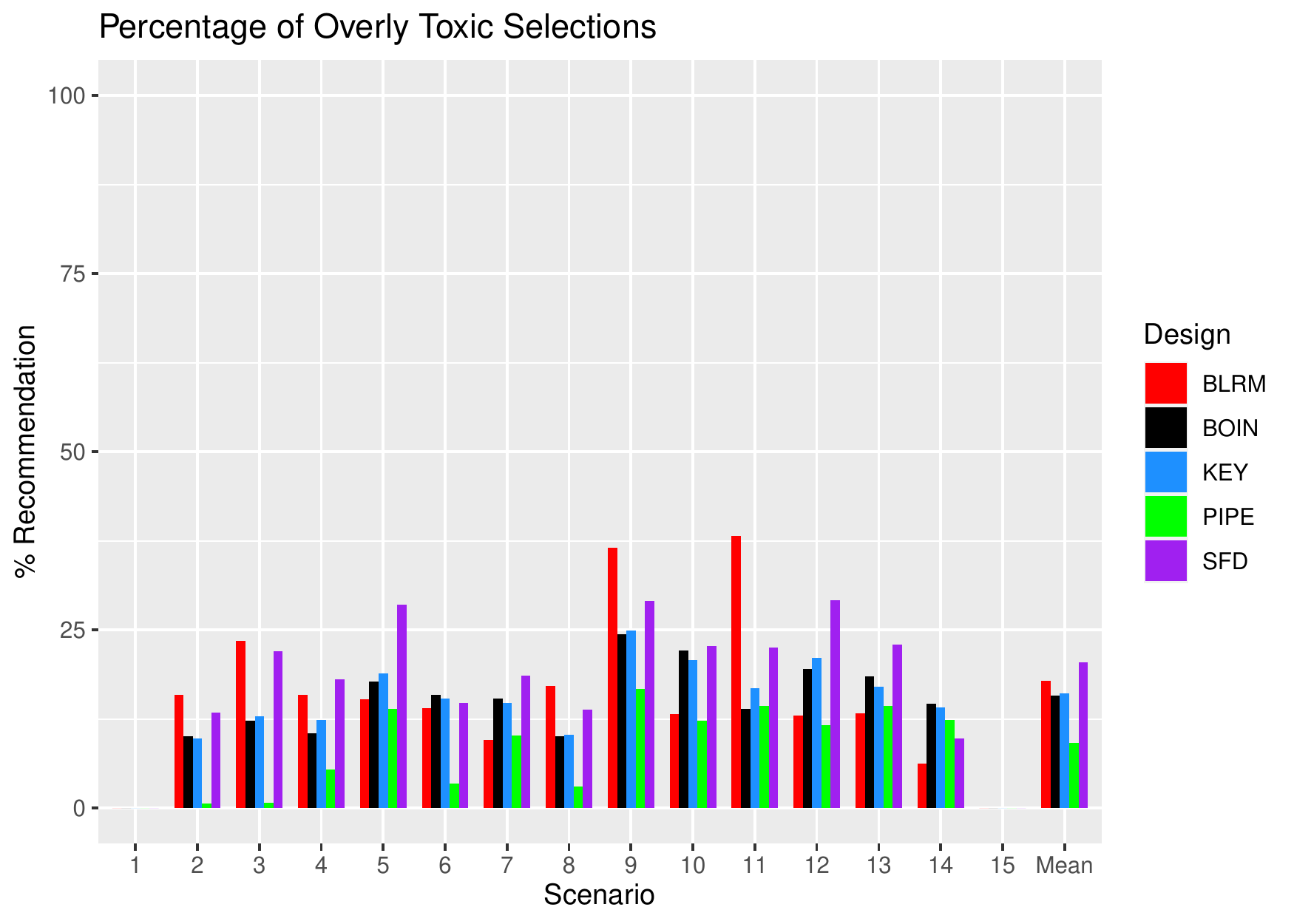}
    \caption{An illustration of the proportion of overly toxic selections across Scenarios 1-15 for each design. The rightmost group of bars show the means.}
    \label{fig:toxsel}
\end{figure}

Figure \ref{fig:toxsel} illustrates the proportion of overly toxic selections for each design. Scenarios 1 and 15 have no overly toxic combinations, so the proportion is zero for these cases. We observe that the SFD and the BLRM recommend more overly toxic combinations on average, in 20.4\% and 17.8\% of trials respectively. This is evidence of the trade-off between selecting combinations close to $\phi$ and the willingness to recommend more overly toxic combinations.

In three of the scenarios, the SFD recommends overly toxic combinations in over 25\% of the simulated trials and in 6 of the scenarios, it is the design with the highest proportion of overly toxic recommendations. The BLRM stands out in Scenarios 9 and 11 with a very high percentage of simulated trials recommending overly toxic doses, driving up the average across scenarios. 

The PIPE design demonstrates a very low proportion of overly toxic selections with a mean of 9.2\% across the 13 scenarios, 6.2\% below any of the other designs. It has the lowest in all but three scenarios. This is a further illustration of the feature of the design to recommend combinations near but lower than the estimated MTC contour.

A focus on Scenario 14, where all dose combinations are overly toxic, shows the BLRM is the most efficient at stopping for safety, with 93.7\% of simulations not recommending any dose combination.

\subsubsection{Number of Patients Treated at Overly Toxic Combinations}
\begin{figure}[ht!]
    \centering
    \includegraphics[width=0.99\textwidth]{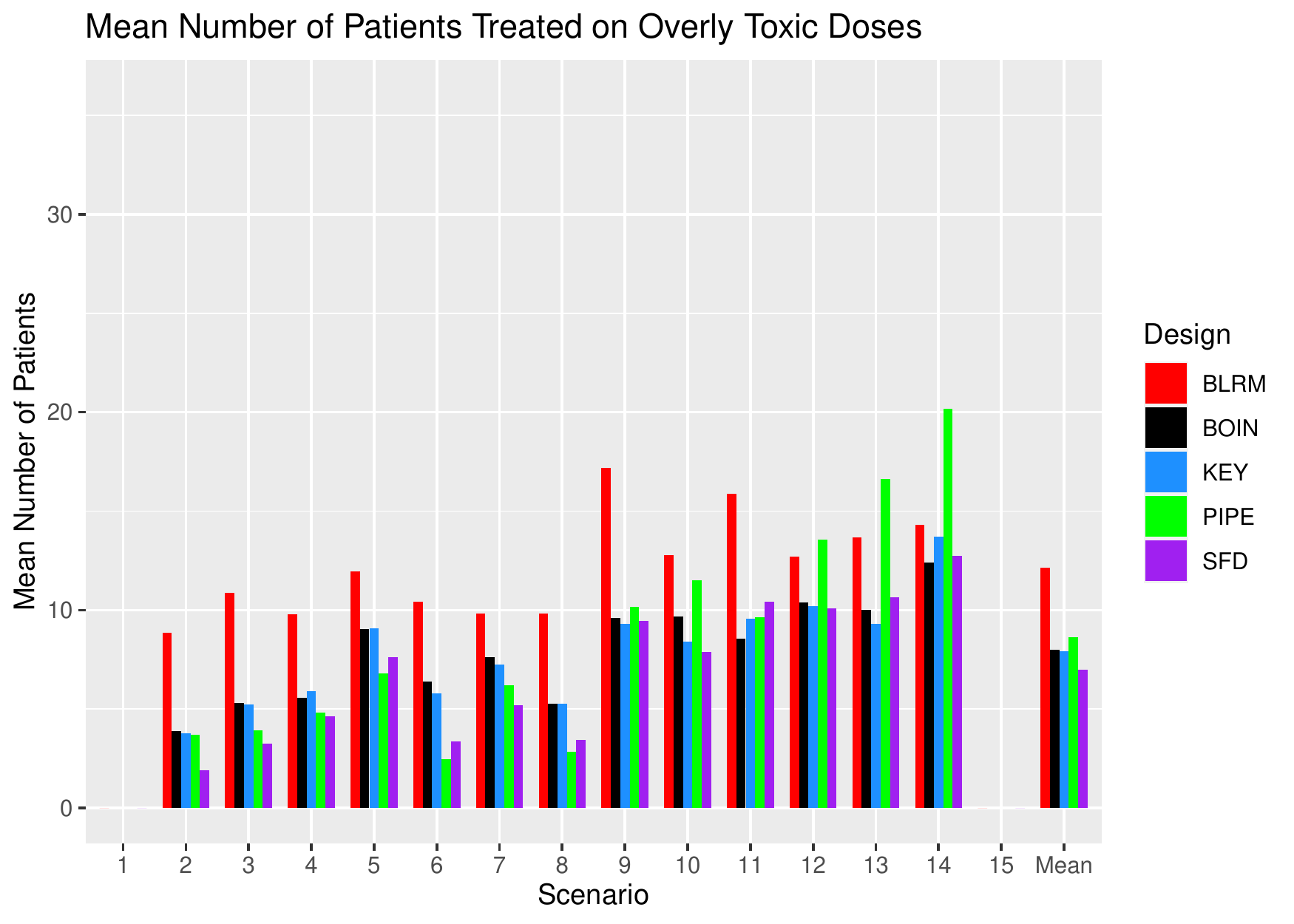}
    \caption{An illustration of the number of patients treated at overly toxic combinations during trials in Scenarios 1-15 for each design. The rightmost group of bars show the means.}
    \label{fig:toxexp}
\end{figure}

Figure \ref{fig:toxexp} outlines the mean number of patients treated at overly toxic combinations in Scenarios 1-15 for each design. Note that we report the number rather than proportion of patients, as this will also give insight into how effectively each design stops for safety.

The most notable feature of these results are the large number of patients treated at overly toxic combinations by the BLRM design. This aggressive escalation is driven by the informative prior, calibrated to give high values of PCS. We refer the reader to the online supplementary materials where an alternative prior leading to more conservative escalation (but considerably lower PCS and PAS) is explored.  

The SFD, KEY and BOIN have reasonable performance, with the SFD showing a strong performance with the lowest overall mean number of patients treated on overly toxic doses of seven patients.

Careful attention must again be paid to Scenario 14, where all dose combinations are overly toxic. The PIPE design treats an average of 20 patients per trial, over six cohorts, which is an unacceptable level of exploration in such a scenario. In this scenario, we also consider that although the BLRM showed good performance in stopping early for safety in the highest number of simulated trials, it also has a high number of patients treated on average before stopping. 

We see that overall the model-free approaches are more conservative in their escalation than the BLRM, with fewer patients treated on unsafe doses, with no noticeable increase in PCS.  Of the model-free approaches, the SFD shows the most promising PCS, at the cost of somewhat higher overly toxic selections. It is also worth noting that the SFD has a substantially higher computational cost than the other model-free designs. To investigate the escalation behaviour further, we consider the application of each method to a case study in the following section.
\section{Case Study} \label{sec:example}

The simulation study gives insight into the operating characteristics of each design, however for further insight into the escalation behaviour, we apply each method to an example case study. We consider a phase I oncology (breast and lung cancer) study enrolling patients to dosing combinations of four dose levels of neratinib and temsirolimus~\citep{Gandhi2014}. A total sample size of 60 patients (cohorts of size 2 or 3) were treated on 12 of 16 possible dosing combinations. Results from 52 patients were included and 10 DLTs were observed, with full results of the trial displayed in Table~\ref{tab:case_study_designs}.

The purpose of this case study is not to investigate whether each design chooses the same MTC as the real study did, but to give an illustration of how each design explores the dosing grid, given identical patient responses.

In order to use the calibrated prior specifications, and in line with the simulation study, we restrict the dosing grid to three doses of each drug, removing the lowest dose of temsirolimus and the highest dose of neratinib. We also fix the cohort size to three patients and maximum total sample size to 36.

To ensure a fair comparison between designs, we define a fixed set of 36 ordered patient responses for each dose combination. The first patient responses in this set are the true $y_{ij}$ DLT responses and $n_{ij}-y_{ij}$ non-DLT responses, in a random permutation (note that this is the same random permutation for each of the methods). The remaining $36-n_{ij}$ responses are generated in the following way. Each patient has an individual probability of DLT, generated from Beta$(1+y_{ij},1+n_{ij}-y_{ij})$. Then a binary response is generated with this probability. Where there were no patients assigned to the dose combination in the real study, the individual P(DLT) is generated from a Beta(3,3) distribution, to indicate the dose combination is unsafe, since this is the reason the combination was not escalated to. This process uses the information from the real study, but also introduces enough variability in the subsequent responses to account for the small sample size. 

Table~\ref{tab:case_study_designs} displays the results of each of the methods, with the number of patients treated at each combination, the number of DLTs observed, and the concluded MTC highlighted in bold. 

The BOIN and KEY designs show very similar exploration, first escalating in neratinib, then temsirolimus. The highest combination is not explored, as the combinations with the next lowest dose of each drug were considered unsafe. The only difference is that the KEY assigns one more cohort to the 200mg/50mg combination, even when the previous cohort had 2/3 observed DLT responses.

The PIPE design explores differently, not escalating to the highest dose of temsirolimus at all, even though only 1/12 DLT responses were observed on the 160mg/50mg combination. The SFD explores more of the highest dose of temsirolimus than any other of the model-free designs, although still not the highest combination. An interesting observation here is that the final recommended dose has observed 4/9 DLT responses, a level that would generally be an unsafe standard. This is in line with the simulation results that showed this design to have the highest level of overly toxic selections.

The BLRM is executed with two prior distributions, the calibrated prior and the alternative prior. Both show a more aggressive escalation than the model free designs, with patients allocated to the highest combination. The calibrated prior gives a slightly more aggressive approach with a second cohort assigned, even when the first observed 2/3 DLT responses. This also means that the dosing grid is not as well explored as some of the model-free designs, as the lowest dose of temsirolimus is only explored in combination with the lowest dose of neratinib. These results are in line with the simulation study for the calibrated prior, where the BLRM had on average the most patients treated on overly toxic doses and also a high proportion of overly toxic recommendations.

The case study highlights some key differences in the approaches, illustrating how both the escalation schemes and final recommendation differ. Particularly of note is the somewhat aggressive behaviour of not de-escalating when observing 2/3 observed DLT responses, and recommending a final dose combination with 4/9 observed DLT responses from  both the SFD and BLRM. This behaviour, that could be considered unsafe, is not necessarily obvious from simulation results and underlines the importance of studying the individual escalations in an example case study. It is also important to consider that in practice, such a statistical approach is a guidance for dose recommendation that should be supported by an overall evaluation of the safety, pharmacokinetics and clinical rationale.

\section{Discussion} \label{sec:discussion}
This paper provides a review of a wide range of combination designs in phase I oncology, exploring the more recently proposed model-free designs in detail, as well as providing a novel approach for the calibration of such designs. The comprehensive simulation study we conduct suggests that model-free designs are competitive with the BLRM in terms of the proportion of correct combinations selected. The operating characteristics of model-free designs in a number of scenarios suggest they offer a safer alternative. The case study example highlighted the key differences in how the methods explore the dosing grid given the same patient responses, with more aggressive approaches missing the lower doses, and conservative approaches missing the higher ones.

The discussed results depend upon the specification of the intervals for the BOIN and KEY designs, and the operational priors for the PIPE, SFD and BLRM designs, which were calibrated using a novel approach. This included calibrating the overdosing rules in each design to reduce the risk of recommending overly toxic combinations for phase II. Naturally, our work does not allow for comparison between designs when complete and reliable prior information on the toxicity of each drug is available. In practice, the PIPE, SFD and BLRM designs can exploit this prior knowledge to help the escalation process. 

The calibration procedure, although novel in approach, is relatively straightforward to implement. It does however highlight the computational intensity of the different methods. Both the BLRM and SFD are very computationally intensive, with the calibration procedure taking substantially longer than for any of the other designs. It has shown great promise in specifying prior distributions that yield high PCS values.

Moreover, our simulations do not allow for the early selection of an MTC. For example, if at least 9 patients are treated at a combination and the next cohort is recommended to be treated at this combination, then a trial could be stopped and this combination selected as the MTC. We acknowledge this rule is useful to reduce sample sizes, especially in scenarios where the true MTC is a low dose combination. Another limitation in the work is in only evaluating $3\times3$ combination grids. This was chosen as a balance between providing a large enough grid to observe interesting differences between designs, but at the same time being computationally feasible and realistic of a dose-finding study and sample size. We hence acknowledge our results may not necessarily hold for all settings with varying grid sizes, and emphasize that the prior specifications we have recommended here for comparison would need to be re-calibrated for a different grid size, sample size or cohort size. 

An additional area of interest for such dose-finding studies is the sample size and cohort size. Conducting a sensitivity analysis on both of these for each design would be an excellent opportunity to investigate whether designs can still achieve high PCS with fewer patients, or significantly higher PCS with extra patients, and whether a larger or smaller cohort size would lead to better exploration of the dosing grid.

Finally, we conclude this comparison with an overview of recommendations for the use of each design in the context of this work. The BOIN and KEY designs give a balanced approach, with a good level of PCS and PAS across a range of scenarios. Overly toxic explorations and selections are also well balanced across scenarios. The PIPE design is more cautious in its selection, with a consistently low proportion of overly toxic selections, although at the cost of also recommending correct combinations a lower proportion of the time. The Surface Free design offers a high PCS and PAS and a generally low number of patients treated at overly toxic selections, but this must be balanced with the high proportion of overly toxic selections. The BLRM provides the most aggressive approach with a calibrated prior, with a large number of patients treated on overly toxic doses, however a good level of PCS and PAS. With an alternative, intuitive prior, the number of overly toxic explorations is reduced, but at the cost of the high PCS values.

\section{Data Availability Statement}
The data that supports the findings of this research are available in the Table~\ref{tab:case_study_designs} of this article, originally from Gandhi et al. \cite{Gandhi2014}, with all other data simulated according to the specifications described.

\section{Acknowledgements}
This report is independent research supported by the National Institute for Health Research (NIHR Advanced Fellowship, Dr Pavel Mozgunov, NIHR300576; and Prof Jaki’s Senior Research Fellowship, NIHR-SRF-2015-08-001) and by the NIHR Cambridge Biomedical Research Centre (BRC-1215-20014). The views expressed in this publication are those of the authors and not necessarily those of the NHS, the National Institute for Health Research or the Department of Health and Social Care (DHSC). T Jaki, P Mozgunov and H Barnett received funding from UK Medical Research Council (MC\_UU\_00002/14).

\bibliographystyle{abbrv}
\bibliography{combination_bib.bib}

\begin{thebibliography}{10}

\bibitem{abbas2020comparison}
R.~Abbas, C.~Rossoni, T.~Jaki, X.~Paoletti, and P.~Mozgunov.
\newblock {A comparison of phase {I} dose-finding designs in clinical trials
  with monotonicity assumption violation}.
\newblock {\em Clinical Trials}, 17(5):522--534, 2020.

\bibitem{conaway2019impact}
M.~R. Conaway and G.~R. Petroni.
\newblock {The impact of early-phase trial design in the drug development
  process}.
\newblock {\em Clinical Cancer Research}, 25(2):819--827, 2019.

\bibitem{couzin2013cancer}
J.~Couzin-Frankel.
\newblock {Cancer Immunotherapy}.
\newblock {\em Science}, 342(6165):1432--1433, 2013.

\bibitem{dykstra1982algorithm}
R.~L. Dykstra and T.~Robertson.
\newblock {An algorithm for isotonic regression for two or more independent
  variables}.
\newblock {\em The Annals of Statistics}, 10(3):708--716, 1982.

\bibitem{Gandhi2014}
L.~Gandhi, R.~Bahleda, S.~M. Tolaney, E.~L. Kwak, J.~M. Cleary, S.~S. Pandya,
  A.~Hollebecque, R.~Abbas, R.~Ananthakrishnan, A.~Berkenblit, M.~Krygowski,
  Y.~Liang, K.~W. Turnbull, G.~I. Shapiro, and J.~C. Soria.
\newblock {Phase I study of neratinib in combination with temsirolimus in
  patients with human epidermal growth factor receptor 2-dependent and other
  solid tumors}.
\newblock {\em Journal of Clinical Oncology}, 32(2):68--75, 2014.

\bibitem{Hamberg2009}
P.~Hamberg and J.~Verweij.
\newblock {Phase I Drug Combination Trial Design: Walking the Tightrope}.
\newblock {\em Journal of Clinical Oncology}, 27(27):4441--4443, 2009.
\newblock PMID: 19704054.

\bibitem{Lin2017boin}
R.~Lin and G.~Yin.
\newblock {Bayesian optimal interval design for dose finding in
  drug-combination trials}.
\newblock {\em Statistical Methods in Medical Research}, 26(5):2155--2167,
  2017.

\bibitem{mander2015product}
A.~P. Mander and M.~J. Sweeting.
\newblock {A product of independent beta probabilities dose escalation design
  for dual-agent phase {I} trials}.
\newblock {\em Statistics in medicine}, 34(8):1261--1276, 2015.

\bibitem{mozgunov2020surface}
P.~Mozgunov, M.~Gasparini, and T.~Jaki.
\newblock {A surface-free design for phase {I} dual-agent combination trials}.
\newblock {\em Statistical Methods in Medical Research}, 29(10):3093--3109,
  2020.

\bibitem{mozgunov2019information}
P.~Mozgunov and T.~Jaki.
\newblock {An information theoretic phase {I}--{II} design for molecularly
  targeted agents that does not require an assumption of monotonicity}.
\newblock {\em Journal of the Royal Statistical Society. Series C, Applied
  Statistics}, 68(2):347, 2019.

\bibitem{mozgunov2020benchmark}
P.~Mozgunov, T.~Jaki, and X.~Paoletti.
\newblock {A benchmark for dose finding studies with continuous outcomes}.
\newblock {\em Biostatistics}, 21(2):189--201, 2020.

\bibitem{mozgunov2021benchmark}
P.~Mozgunov, X.~Paoletti, and T.~Jaki.
\newblock {A benchmark for dose-finding studies with unknown ordering}.
\newblock {\em Biostatistics}, 2021.

\bibitem{neuenschwander2015bayesian}
B.~Neuenschwander, A.~Matano, Z.~Tang, S.~Roychoudhury, S.~Wandel, and
  S.~Bailey.
\newblock {A Bayesian industry approach to phase {I} combination trials in
  oncology}.
\newblock In {\em Statistical methods in drug combination studies}, pages
  95--135. Chapman \& Hall/CRC Press: Boca Raton, FL, 2015.

\bibitem{R}
{R Core Team}.
\newblock {\em {R: A Language and Environment for Statistical Computing}}.
\newblock R Foundation for Statistical Computing, Vienna, Austria, 2020.

\bibitem{riviere2015competing}
M.-K. Riviere, F.~Dubois, and S.~Zohar.
\newblock {Competing designs for drug combination in phase {I} dose-finding
  clinical trials}.
\newblock {\em Statistics in medicine}, 34(1):1--12, 2015.

\bibitem{riviere2015designs}
M.-K. Riviere, C.~Le~Tourneau, X.~Paoletti, F.~Dubois, and S.~Zohar.
\newblock {Designs of drug-combination phase I trials in oncology: a systematic
  review of the literature}.
\newblock {\em Annals of Oncology}, 26(4):669--674, 2015.

\bibitem{sharma2015immune}
P.~Sharma and J.~P. Allison.
\newblock {Immune checkpoint targeting in cancer therapy: toward combination
  strategies with curative potential}.
\newblock {\em Cell}, 161(2):205--214, 2015.

\bibitem{thall2003practical}
P.~F. Thall and S.-J. Lee.
\newblock {Practical model-based dose-finding in phase I clinical trials:
  methods based on toxicity}.
\newblock {\em International Journal of Gynecologic Cancer}, 13(3):251--261,
  2003.

\bibitem{thall2003dose}
P.~F. Thall, R.~E. Millikan, P.~Mueller, and S.-J. Lee.
\newblock {Dose-finding with two agents in phase I oncology trials}.
\newblock {\em Biometrics}, 59(3):487--496, 2003.

\bibitem{wages2016practical}
N.~A. Wages, A.~Ivanova, and O.~Marchenko.
\newblock {Practical designs for phase I combination studies in oncology}.
\newblock {\em Journal of Biopharmaceutical Statistics}, 26(1):150--166, 2016.

\bibitem{wong2019estimation}
C.~H. Wong, K.~W. Siah, and A.~W. Lo.
\newblock {Estimation of clinical trial success rates and related parameters}.
\newblock {\em Biostatistics}, 20(2):273--286, 2019.

\bibitem{wong2016changing}
K.~M. Wong, A.~Capasso, and S.~G. Eckhardt.
\newblock {The changing landscape of phase I trials in oncology}.
\newblock {\em Nature Reviews Clinical oncology}, 13(2):106--117, 2016.

\bibitem{yan2017keyboard}
F.~Yan, S.~J. Mandrekar, and Y.~Yuan.
\newblock {Keyboard: a novel Bayesian toxicity probability interval design for
  phase {I} clinical trials}.
\newblock {\em Clinical Cancer Research}, 23(15):3994--4003, 2017.

\end{thebibliography}

\bgroup
\def\arraystretch{1.2}
\begin{table}[p!]
\begin{minipage}{0.3\textwidth}
\centering
\begin{tabular}{|c|ccc|} 
\multicolumn{4}{c}{\textbf{Scenario 1}} \vspace{2mm} \\ 
\hline
& $d_1^B$ & $d_2^B$ & $d_3^B$ \\ 
\hline
$d_1^A$ & 0.05 & 0.10 & 0.15 \\
$d_2^A$ & 0.10 & 0.15 & \uu{0.20} \\ 
$d_3^A$ & 0.15 & \uu{0.20} & \uu{\textbf{0.30}} \\ 
\hline
\end{tabular}
\end{minipage}
\begin{minipage}{0.38\textwidth}
\centering
\begin{tabular}{|c|ccc|}
\multicolumn{4}{c}{\textbf{Scenario 2}} \vspace{2mm} \\ 
\hline
& $d_1^B$ & $d_2^B$ & $d_3^B$ \\ 
\hline
$d_1^A$ & 0.05 & 0.10 & 0.15 \\
$d_2^A$ & 0.10 & \uu{0.20} & \uu{\textbf{0.30}} \\ 
$d_3^A$ & \uu{0.20} & \uu{\textbf{0.30}} & 0.45 \\ 
\hline
\end{tabular}
\end{minipage}
\begin{minipage}{0.3\textwidth}
\centering
\begin{tabular}{|c|ccc|}
\multicolumn{4}{c}{\textbf{Scenario 3}} \vspace{2mm} \\ 
\hline
& $d_1^B$ & $d_2^B$ & $d_3^B$ \\ 
\hline
$d_1^A$ & 0.02 & 0.05 & 0.10 \\
$d_2^A$ & 0.10 & 0.15 & \uu{0.20} \\ 
$d_3^A$ & \uu{0.20} & \uu{\textbf{0.30}} & 0.45 \\ 
\hline
\end{tabular}
\end{minipage}

\vspace{6mm}

\begin{minipage}{0.3\textwidth}
\centering
\begin{tabular}{|c|ccc|}
\multicolumn{4}{c}{\textbf{Scenario 4}} \vspace{2mm} \\ 
\hline
& $d_1^B$ & $d_2^B$ & $d_3^B$ \\ 
\hline
$d_1^A$ & 0.05 & 0.10 & 0.15 \\
$d_2^A$ & 0.10 & \uu{0.20} & \uu{\textbf{0.30}} \\ 
$d_3^A$ & \uu{0.20} & 0.45 & 0.60 \\ 
\hline
\end{tabular}
\end{minipage}
\begin{minipage}{0.38\textwidth}
\centering
\begin{tabular}{|c|ccc|}
\multicolumn{4}{c}{\textbf{Scenario 5}} \vspace{2mm} \\ 
\hline
& $d_1^B$ & $d_2^B$ & $d_3^B$ \\ 
\hline
$d_1^A$ & 0.02 & 0.05 & 0.15 \\
$d_2^A$ & \uu{0.20} & \uu{\textbf{0.30}} & 0.45 \\ 
$d_3^A$ & 0.45 & 0.55 & 0.65 \\ 
\hline
\end{tabular}
\end{minipage}
\begin{minipage}{0.3\textwidth}
\centering
\begin{tabular}{|c|ccc|}
\multicolumn{4}{c}{\textbf{Scenario 6}} \vspace{2mm} \\ 
\hline
& $d_1^B$ & $d_2^B$ & $d_3^B$ \\ 
\hline
$d_1^A$ & 0.10 & 0.15 & \uu{\textbf{0.30}} \\
$d_2^A$ & 0.15 & \uu{\textbf{0.30}} & 0.45 \\ 
$d_3^A$ & \uu{\textbf{0.30}} & 0.45 & 0.60 \\ 
\hline
\end{tabular}
\end{minipage}

\vspace{6mm}

\begin{minipage}{0.3\textwidth}
\centering
\begin{tabular}{|c|ccc|}
\multicolumn{4}{c}{\textbf{Scenario 7}} \vspace{2mm} \\ 
\hline
& $d_1^B$ & $d_2^B$ & $d_3^B$ \\ 
\hline
$d_1^A$ & 0.10 & \uu{0.20} & 0.45 \\
$d_2^A$ & 0.15 & \uu{\textbf{0.30}} & 0.50 \\ 
$d_3^A$ & \uu{\textbf{0.30}} & 0.50 & 0.60 \\ 
\hline
\end{tabular}
\end{minipage}
\begin{minipage}{0.38\textwidth}
\centering
\begin{tabular}{|c|ccc|}
\multicolumn{4}{c}{\textbf{Scenario 8}} \vspace{2mm} \\ 
\hline
& $d_1^B$ & $d_2^B$ & $d_3^B$ \\ 
\hline
$d_1^A$ & 0.05 & 0.10 & \uu{0.20} \\
$d_2^A$ & 0.10 & \uu{0.20} & \uu{\textbf{0.30}} \\ 
$d_3^A$ & \uu{\textbf{0.30}} & 0.45 & 0.55 \\ 
\hline
\end{tabular}
\end{minipage}
\begin{minipage}{0.3\textwidth}
\centering
\begin{tabular}{|c|ccc|}
\multicolumn{4}{c}{\textbf{Scenario 9}} \vspace{2mm} \\ 
\hline
& $d_1^B$ & $d_2^B$ & $d_3^B$ \\ 
\hline
$d_1^A$ & 0.10 & 0.15 & \uu{\textbf{0.30}} \\
$d_2^A$ & \uu{\textbf{0.30}} & 0.40 & 0.50 \\ 
$d_3^A$ & 0.40 & 0.50 & 0.60 \\ 
\hline
\end{tabular}
\end{minipage}

\vspace{6mm}

\begin{minipage}{0.3\textwidth}
\centering
\begin{tabular}{|c|ccc|}
\multicolumn{4}{c}{\textbf{Scenario 10}} \vspace{2mm} \\ 
\hline
& $d_1^B$ & $d_2^B$ & $d_3^B$ \\ 
\hline
$d_1^A$ & 0.15 & \uu{\textbf{0.30}} & 0.45 \\
$d_2^A$ & \uu{\textbf{0.30}} & 0.45 & 0.55 \\ 
$d_3^A$ & 0.45 & 0.55 & 0.65 \\ 
\hline
\end{tabular}
\end{minipage}
\begin{minipage}{0.38\textwidth}
\centering
\begin{tabular}{|c|ccc|}
\multicolumn{4}{c}{\textbf{Scenario 11}} \vspace{2mm} \\ 
\hline
& $d_1^B$ & $d_2^B$ & $d_3^B$ \\ 
\hline
$d_1^A$ & 0.02 & 0.05 & 0.10 \\
$d_2^A$ & \uu{\textbf{0.30}} & 0.45 & 0.60 \\ 
$d_3^A$ & 0.45 & 0.60 & 0.75 \\ 
\hline
\end{tabular}
\end{minipage}
\begin{minipage}{0.3\textwidth}
\centering
\begin{tabular}{|c|ccc|}
\multicolumn{4}{c}{\textbf{Scenario 12}} \vspace{2mm} \\ 
\hline
& $d_1^B$ & $d_2^B$ & $d_3^B$ \\ 
\hline
$d_1^A$ & \uu{0.20} & \uu{\textbf{0.30}} & 0.45 \\
$d_2^A$ & 0.45 & 0.50 & 0.55 \\ 
$d_3^A$ & 0.65 & 0.70 & 0.75 \\ 
\hline
\end{tabular}
\end{minipage}

\vspace{5mm}

\begin{minipage}{0.3\textwidth}
\centering
\begin{tabular}{|c|ccc|}
\multicolumn{4}{c}{\textbf{Scenario 13}} \vspace{2mm} \\ 
\hline
& $d_1^B$ & $d_2^B$ & $d_3^B$ \\ 
\hline
$d_1^A$ & \uu{\textbf{0.30}} & 0.45 & 0.50 \\
$d_2^A$ & 0.45 & 0.50 & 0.55 \\ 
$d_3^A$ & 0.50 & 0.55 & 0.60 \\ 
\hline
\end{tabular}
\end{minipage}
\begin{minipage}{0.38\textwidth}
\centering
\begin{tabular}{|c|ccc|}
\multicolumn{4}{c}{\textbf{Scenario 14}} \vspace{2mm} \\ 
\hline
& $d_1^B$ & $d_2^B$ & $d_3^B$ \\ 
\hline
$d_1^A$ & 0.45 & 0.50 & 0.55 \\
$d_2^A$ & 0.50 & 0.55 & 0.60 \\ 
$d_3^A$ & 0.55 & 0.60 & 0.65 \\ 
\hline
\end{tabular}
\end{minipage}
\begin{minipage}{0.3\textwidth}
\centering
\begin{tabular}{|c|ccc|}
\multicolumn{4}{c}{\textbf{Scenario 15}} \vspace{2mm} \\ 
\hline
& $d_1^B$ & $d_2^B$ & $d_3^B$ \\ 
\hline
$d_1^A$ & 0.10 & 0.10 & 0.10 \\
$d_2^A$ & 0.10 & 0.10 & 0.10 \\
$d_3^A$ & 0.10 & 0.10 & 0.10 \\ 
\hline
\end{tabular}
\end{minipage}
\vspace{4mm}
\caption{Toxicity scenarios to evaluate the combination designs. Rows and columns refer to the dose of drug A and B respectively. True MTCs are in bold and `acceptable' combinations are underlined. \label{tab:scenarios}}
\end{table}
\egroup

\bgroup
\def\arraystretch{1.2}

\begin{table}[p!]
\begin{minipage}{0.99\textwidth}
\centering
\begin{tabular}{cccccc}
&Raw Trial Data &\multicolumn{4}{c}{\textbf{Temsirolimus}}  \\
  \cline{2-6}
 &  & 15mg & 25mg & 50mg & 75mg \\ 
 \cline{2-6}
 & 120mg & 0/2 & 0/4 & 1/5 & 0/4 \\ 
 \textbf{Neratinib} & 160mg & 1/4 & 1/4 & \textbf{0/5} & 3/6 \\ 
   & 200mg & 0/4 & \textbf{1/8}& 1/2 &  \\ 
   & 240mg & 2/4 &  &  &  \\ 
   \cline{2-6}
\end{tabular}
\end{minipage}
\vspace{5mm}

\begin{minipage}{0.49\textwidth}
\centering
\begin{tabular}{ccccc}
&\hspace{4mm}BOIN\hspace{4mm}&\multicolumn{3}{c}{\textbf{Temsirolimus}}  \\
  \cline{2-5}
 &  & 25mg & 50mg & 75mg \\ 
 \cline{2-5}
    & 120mg & 0/3 & 0/0 & 0/0 \\ 
  \textbf{Neratinib} & 160mg & 1/6 & 0/6 & 3/6 \\ 
   & 200mg & \textbf{2/12} & 2/3 & 0/0 \\ 
   
   \cline{2-5}
\end{tabular}
\end{minipage}
\begin{minipage}{0.49\textwidth}
\centering
\begin{tabular}{ccccc}
& KEY &\multicolumn{3}{c}{\textbf{Temsirolimus}}  \\
  \cline{2-5}
 &  & 25mg & 50mg & 75mg \\ 
 \cline{2-5}
 & 120mg & 0/3 & 0/0 & 0/0 \\ 
  \textbf{Neratinib} & 160mg & 1/6 & 0/6 & 3/6 \\ 
   & 200mg & \textbf{2/9} & 3/6 & 0/0 \\ 
   
   \cline{2-5}
\end{tabular}
\end{minipage}

\vspace{6mm}

\begin{minipage}{0.49\textwidth}
\centering
\begin{tabular}{ccccc}
&\hspace{5mm}PIPE\hspace{5mm}&\multicolumn{3}{c}{\textbf{Temsirolimus}}  \\
  \cline{2-5}
 &  & 25mg & 50mg & 75mg \\ 
 \cline{2-5}
  & 120mg & 0/3 & 0/0 & 0/0 \\ 
  \textbf{Neratinib} & 160mg & 1/3 & \textbf{1/12} & 0/0 \\ 
   & 200mg & \textbf{2/15} & 2/3 & 0/0 \\ 
   \cline{2-5}
\end{tabular}
\end{minipage}
\begin{minipage}{0.49\textwidth}
\centering
\begin{tabular}{ccccc}
&\hspace{10mm}SFD\hspace{10mm}&\multicolumn{3}{c}{\textbf{Temsirolimus}}  \\
  \cline{2-5}
 &  & 25mg & 50mg & 75mg \\ 
 \cline{2-5}
  & 120mg & 0/3 & 0/0 & 0/6 \\ 
  \textbf{Neratinib} & 160mg & 1/6 & 0/6 & \textbf{4/9} \\ 
   & 200mg & 0/3 & 2/3 & 0/0 \\
   \cline{2-5}
\end{tabular}
\end{minipage}

\vspace{6mm}

\begin{minipage}{0.49\textwidth}
\centering
\begin{tabular}{ccccc}
&BLRM (c)&\multicolumn{3}{c}{\textbf{Temsirolimus}}  \\
  \cline{2-5}
 &  & 25mg & 50mg & 75mg \\ 
 \cline{2-5}
 & 120mg & 0/3 & 0/3 & 0/6 \\ 
  \textbf{Neratinib} & 160mg & 0/0 & 0/9 & \textbf{4/9}\\ 
   & 200mg & 0/0 & 0/0 & 5/6 \\ 
   \cline{2-5}
\end{tabular}
\end{minipage} 
\begin{minipage}{0.49\textwidth}
\centering
\begin{tabular}{ccccc}
&BLRM (a)&\multicolumn{3}{c}{\textbf{Temsirolimus}}  \\
  \cline{2-5}
 &  & 25mg & 50mg & 75mg \\ 
 \cline{2-5}
   & 120mg & 0/3 & 0/3 & 0/6 \\ 
  \textbf{Neratinib} & 160mg & 0/0 & 0/9 & \textbf{4/12} \\ 
   & 200mg & 0/0 & 0/0 & 2/3 \\ 
   \cline{2-5}
\end{tabular}
\end{minipage}
\vspace{6mm}

\caption{Results for each of the designs applied to the case study, including the raw trial data of the study by Gandhi et al. \cite{Gandhi2014}. Each entry represents $y_{ij}/n_{ij}$. The MTC as chosen by each design is highlighed in bold. In the case of the BLRM, (c) indicates the calibrated prior hyperparameters were used and (a) indicates the alternative values were used. \label{tab:case_study_designs}}
\end{table}
\egroup

\end{document}